%% file: Mrk501.tex
\documentclass[iop]{emulateapj}

\usepackage{graphicx}
\usepackage{amssymb,amsmath}
\usepackage[dvips]{color}
\usepackage{ulem}

\slugcomment{Accepted by ApJ}

\shorttitle{Mrk 501}
\shortauthors{Furniss et al.}

\begin{document}


\title{First \textit{NuSTAR} Observations of Mrk\,501 within a Radio to TeV Multi-Instrument Campaign}

\author{A.~Furniss\altaffilmark{1}, 
K.~Noda\altaffilmark{2}, S.~Boggs\altaffilmark{3}, J.~Chiang\altaffilmark{4}, F.~Christensen\altaffilmark{5}, W.~Craig\altaffilmark{6}, P~. Giommi\altaffilmark{7}, C.~Hailey\altaffilmark{8}, F.~Harisson\altaffilmark{9}, G. Madejski\altaffilmark{4}, K.~Nalewajko\altaffilmark{4}
% ...
\\
(The MAGIC Collaboration),\\ 
A.~Archer\altaffilmark{45},
W.~Benbow\altaffilmark{46},
R.~Bird\altaffilmark{47},
J.~Biteau\altaffilmark{48},
V.~Bugaev\altaffilmark{45},
J.~V~Cardenzana\altaffilmark{49},
M.~Cerruti\altaffilmark{46},
X.~Chen\altaffilmark{50,51},
L.~Ciupik\altaffilmark{52},
M.~P.~Connolly\altaffilmark{53},
W.~Cui\altaffilmark{54},
H.~J.~Dickinson\altaffilmark{49},
J.~Dumm\altaffilmark{55},
J.~D.~Eisch\altaffilmark{49},
A.~Falcone\altaffilmark{56},
Q.~Feng\altaffilmark{54},
J.~P.~Finley\altaffilmark{54},
H.~Fleischhack\altaffilmark{51},
P.~Fortin\altaffilmark{46},
L.~Fortson\altaffilmark{55},
L.~Gerard\altaffilmark{51},
G.~H.~Gillanders\altaffilmark{53},
S.~Griffin\altaffilmark{62},
S.~T.~Griffiths\altaffilmark{57},
J.~Grube\altaffilmark{52},
G.~Gyuk\altaffilmark{52},
N.~H{\aa}kansson\altaffilmark{50},
J.~Holder\altaffilmark{58},
T.~B.~Humensky\altaffilmark{59},
C.~A.~Johnson\altaffilmark{48},
P.~Kaaret\altaffilmark{57},
M.~Kertzman\altaffilmark{60},
D.~Kieda\altaffilmark{61},
M.~Krause\altaffilmark{51},
F.~Krennrich\altaffilmark{49},
M.~J.~Lang\altaffilmark{53},
T.~T.Y.~Lin\altaffilmark{62},
G.~Maier\altaffilmark{51},
S.~McArthur\altaffilmark{63},
A.~McCann\altaffilmark{64},
K.~Meagher\altaffilmark{65},
P.~Moriarty\altaffilmark{53},
R.~Mukherjee\altaffilmark{66},
D.~Nieto\altaffilmark{59},
A.~O'Faol\'{a}in de Bhr\'{o}ithe\altaffilmark{51},
R.~A.~Ong\altaffilmark{67},
N.~Park\altaffilmark{63},
D.~Petry\altaffilmark{95},
M.~Pohl\altaffilmark{50,51},
A.~Popkow\altaffilmark{67},
K.~Ragan\altaffilmark{62},
G.~Ratliff\altaffilmark{52},
L.~C.~Reyes\altaffilmark{68},
P.~T.~Reynolds\altaffilmark{69},
G.~T.~Richards\altaffilmark{65},
E.~Roache\altaffilmark{46},
M.~Santander\altaffilmark{66},
G.~H.~Sembroski\altaffilmark{54},
K.~Shahinyan\altaffilmark{55},
D.~Staszak\altaffilmark{62},
I.~Telezhinsky\altaffilmark{50,51},
J.~V.~Tucci\altaffilmark{54},
J.~Tyler\altaffilmark{62},
V.~V.~Vassiliev\altaffilmark{67},
S.~P.~Wakely\altaffilmark{63},
O.~M.~Weiner\altaffilmark{59},
A.~Weinstein\altaffilmark{49},
A.~Wilhelm\altaffilmark{50,51},
D.~A.~Williams\altaffilmark{48},
B.~Zitzer\altaffilmark{70}\\
(The VERITAS Collaboration),\\ 
O.~Vince\altaffilmark{76}\\
L. Fuhrmann\altaffilmark{71},
E. Angelakis \altaffilmark{71},
V. Karamanavis\altaffilmark{71},
I. Myserlis\altaffilmark{71},
T. P. Krichbaum\altaffilmark{71},
J. A. Zensus\altaffilmark{72},
H. Ungerechts\altaffilmark{72},
A. Sievers\altaffilmark{72} \\
R.~Bachev\altaffilmark{73},
M.~B\"ottcher\altaffilmark{74},
W.~P.~Chen\altaffilmark{75},
G.~Damljanovic\altaffilmark{76},
C.~Eswaraiah\altaffilmark{75},
T.~G\"uver\altaffilmark{77},
T.~Hovatta\altaffilmark{9,78},
Z.~Hughes\altaffilmark{48},
S.~.I.~Ibryamov\altaffilmark{79},
M.~D.~Joner\altaffilmark{80},
B.~Jordan\altaffilmark{81},
S.~G.~Jorstad\altaffilmark{82,83},
M.~Joshi\altaffilmark{82},
J.~Kataoka\altaffilmark{84},
O.~M.~Kurtanidze\altaffilmark{85,86},
S.~O.~Kurtanidze\altaffilmark{85},
A.~L\"ahteenm\"aki\altaffilmark{87,88},
G.~Latev\altaffilmark{89},
H.~C.~Lin\altaffilmark{75},
V.~M.~Larionov\altaffilmark{90,91,92},
A.~A.~Mokrushina\altaffilmark{90,91},
D.~A.~Morozova\altaffilmark{90},
M.~G.~Nikolashvili\altaffilmark{85},
C.~M.~Raiteri\altaffilmark{93},
V.~Ramakrishnan\altaffilmark{87},
A.~C.~R. Readhead\altaffilmark{8},
A.~C.~Sadun\altaffilmark{94},
L.~A.~Sigua\altaffilmark{85},
E.~H.~Semkov\altaffilmark{79},
A.~Strigachev\altaffilmark{73},
J.~Tammi\altaffilmark{87},
M. Tornikoski\altaffilmark{87}, 
Y.~V.~Troitskaya\altaffilmark{90},
I.~S.~Troitsky\altaffilmark{90},
M. Villata\altaffilmark{93}
}
\email{amy.furniss@gmail.com}
\email{nodak5@gmail.com}
\email{josefa.becerra@nasa.gov}

\altaffiltext{1}{Department of Physics, Stanford University, Stanford, CA 94305, USA}
\altaffiltext{2} {Max-Planck-Institut f\"ur Physik, D-80805 M\"unchen, Germany}
\altaffiltext{3}{Space Science Laboratory, University of California, Berkeley, CA 94720, USA}
\altaffiltext{4}{Kavli Institute for Particle Astrophysics and Cosmology, SLAC National Accelerator Laboratory, Stanford University, Stanford, CA 94305, USA}
\altaffiltext{5}{DTU Space, National Space Institute, Technical University of Denmark, Elektrovej 327, DK - 2800 Lyngby, Denmark}
\altaffiltext{6}{Lawrence Livermore National Laboratory, Livermore, CA 94550, USA and Space Science Laboratory, University of California, Berkeley, CA 94720, USA}
\altaffiltext{7}{ASI Science Data Center (ASDC) , Italian Space Agency (ASI) Via del Politecnico snc, Rome, Italy}
\altaffiltext{8}{Columbia Astrophysics Laboratory, Columbia University, New York, NY 10027, USA}
\altaffiltext{9}{Cahill Center for Astronomy and Astrophysics, Caltech, Pasadena, CA 91125, USA}
\altaffiltext{10}{INAF-OAR, Via Frascati 33, I00040 Monte Porzio Catone (RM), Italy}
\altaffiltext{11}{Jet Propulsion Laboratory, California Institute of Technology, Pasadena, CA 91109, USA}
\altaffiltext{12}{Yale Center for Astronomy and Astrophysics, Physics Department, Yale University, PO Box 208120, New Haven, CT 06520-8120, USA}
\altaffiltext{13}{NASA Goddard Space Flight Center, Greenbelt, MD 20771, USA}
\altaffiltext{14} {ETH Zurich, CH-8093 Zurich, Switzerland}
\altaffiltext{15} {Universit\`a di Udine, and INFN Trieste, I-33100 Udine, Italy}
\altaffiltext{16} {INAF National Institute for Astrophysics, I-00136 Rome, Italy}
\altaffiltext{17} {Universit\`a  di Siena, and INFN Pisa, I-53100 Siena, Italy}
\altaffiltext{18} {Croatian MAGIC Consortium, Rudjer Boskovic Institute, University of Rijeka and University of Split, HR-10000 Zagreb, Croatia}
\altaffiltext{19} {Saha Institute of Nuclear Physics, 1\textbackslash{}AF Bidhannagar, Salt Lake, Sector-1, Kolkata 700064, India}
\altaffiltext{20} {Universidad Complutense, E-28040 Madrid, Spain}
\altaffiltext{21} {Inst. de Astrof\'isica de Canarias, E-38200 La Laguna, Tenerife, Spain}
\altaffiltext{22} {University of \L\'od\'z, PL-90236 Lodz, Poland}
\altaffiltext{23} {Deutsches Elektronen-Synchrotron (DESY), D-15738 Zeuthen, Germany}
\altaffiltext{24} {IFAE, Campus UAB, E-08193 Bellaterra, Spain}
\altaffiltext{25} {Universit\"at W\"urzburg, D-97074 W\"urzburg, Germany}
\altaffiltext{26} {Centro de Investigaciones Energ\'eticas, Medioambientales y Tecnol\'ogicas, E-28040 Madrid, Spain}
\altaffiltext{27} {Universit\`a di Padova and INFN, I-35131 Padova, Italy}
\altaffiltext{28} {Institute of Space Sciences, E-08193 Barcelona, Spain}
\altaffiltext{29} {Technische Universit\"at Dortmund, D-44221 Dortmund, Germany}
\altaffiltext{30} {Unitat de F\'isica de les Radiacions, Departament de F\'isica, and CERES-IEEC, Universitat Aut\`onoma de Barcelona, E-08193 Bellaterra, Spain}
\altaffiltext{31} {Universitat de Barcelona, ICC, IEEC-UB, E-08028 Barcelona, Spain}
\altaffiltext{32} {Japanese MAGIC Consortium, ICRR, The University of Tokyo, Department of Physics and Hakubi Center, Kyoto University, Tokai University, The University of Tokushima, KEK, Japan}
\altaffiltext{33} {Finnish MAGIC Consortium, Tuorla Observatory, University of Turku and Department of Physics, University of Oulu, Finland}
\altaffiltext{34} {Inst. for Nucl. Research and Nucl. Energy, BG-1784 Sofia, Bulgaria}
\altaffiltext{35} {Universit\`a di Pisa, and INFN Pisa, I-56126 Pisa, Italy}
\altaffiltext{36} {ICREA and Institute of Space Sciences, E-08193 Barcelona, Spain}
\altaffiltext{37} {Universit\`a dell'Insubria and INFN Milano Bicocca, Como, I-22100 Como, Italy}
\altaffiltext{38} {now at Centro Brasileiro de Pesquisas F\'isicas (CBPF\textbackslash{}MCTI), R. Dr. Xavier Sigaud, 150 - Urca, Rio de Janeiro - RJ, 22290-180, Brazil}
\altaffiltext{39} {now at NASA Goddard Space Flight Center, Greenbelt, MD 20771, USA and Department of Physics and Department of Astronomy, University of Maryland, College Park, MD 20742, USA}
\altaffiltext{40} {Humboldt University of Berlin, Istitut f\"ur Physik  Newtonstr. 15, 12489 Berlin Germany}
\altaffiltext{41} {now at Ecole polytechnique f\'ed\'erale de Lausanne (EPFL), Lausanne, Switzerland}
\altaffiltext{42} {now at Finnish Centre for Astronomy with ESO (FINCA), Turku, Finland}
\altaffiltext{43} {also at INAF-Trieste}
\altaffiltext{44} {also at ISDC - Science Data Center for Astrophysics, 1290, Versoix (Geneva)}
\altaffiltext{45}{Department of Physics, Washington University, St. Louis, MO 63130, USA}
\altaffiltext{46}{Fred Lawrence Whipple Observatory, Harvard-Smithsonian Center for Astrophysics, Amado, AZ 85645, USA}
\altaffiltext{47}{School of Physics, University College Dublin, Belfield, Dublin 4, Ireland}
\altaffiltext{48}{Santa Cruz Institute for Particle Physics and Department of Physics, University of California, Santa Cruz, CA 95064, USA}
\altaffiltext{49}{Department of Physics and Astronomy, Iowa State University, Ames, IA 50011, USA}
\altaffiltext{50}{Institute of Physics and Astronomy, University of Potsdam, 14476 Potsdam-Golm, Germany}
\altaffiltext{51}{DESY, Platanenallee 6, 15738 Zeuthen, Germany}
\altaffiltext{52}{Astronomy Department, Adler Planetarium and Astronomy Museum, Chicago, IL 60605, USA}
\altaffiltext{53}{School of Physics, National University of Ireland Galway, University Road, Galway, Ireland}
\altaffiltext{54}{Department of Physics and Astronomy, Purdue University, West Lafayette, IN 47907, USA}
\altaffiltext{55}{School of Physics and Astronomy, University of Minnesota, Minneapolis, MN 55455, USA}
\altaffiltext{56}{Department of Astronomy and Astrophysics, 525 Davey Lab, Pennsylvania State University, University Park, PA 16802, USA}
\altaffiltext{57}{Department of Physics and Astronomy, University of Iowa, Van Allen Hall, Iowa City, IA 52242, USA}
\altaffiltext{58}{Department of Physics and Astronomy and the Bartol Research Institute, University of Delaware, Newark, DE 19716, USA}
\altaffiltext{59}{Physics Department, Columbia University, New York, NY 10027, USA}
\altaffiltext{60}{Department of Physics and Astronomy, DePauw University, Greencastle, IN 46135-0037, USA}
\altaffiltext{61}{Department of Physics and Astronomy, University of Utah, Salt Lake City, UT 84112, USA}
\altaffiltext{62}{Physics Department, McGill University, Montreal, QC H3A 2T8, Canada}
\altaffiltext{63}{Enrico Fermi Institute, University of Chicago, Chicago, IL 60637, USA}
\altaffiltext{64}{Kavli Institute for Cosmological Physics, University of Chicago, Chicago, IL 60637, USA}
\altaffiltext{65}{School of Physics and Center for Relativistic Astrophysics, Georgia Institute of Technology, 837 State Street NW, Atlanta, GA 30332-0430}
\altaffiltext{66}{Department of Physics and Astronomy, Barnard College, Columbia University, NY 10027, USA}
\altaffiltext{67}{Department of Physics and Astronomy, University of California, Los Angeles, CA 90095, USA}
\altaffiltext{68}{Physics Department, California Polytechnic State University, San Luis Obispo, CA 94307, USA}
\altaffiltext{69}{Department of Applied Science, Cork Institute of Technology, Bishopstown, Cork, Ireland}
\altaffiltext{70}{Argonne National Laboratory, 9700 S. Cass Avenue, Argonne, IL 60439, USA}
\altaffiltext{71}{Max-Planck-Institut f\"ur Radioastronomie, Auf dem Huegel 69, 53121 Bonn, Germany}
\altaffiltext{72}{Institut de Radio Astronomie Millim\'etrique, Avenida Divina Pastora 7, Local 20, 18012 Granada, Spain}
\altaffiltext{73}{Institute of Astronomy, Bulgarian Academy of Sciences, 72 Tsarigradsko shosse Blvd., 1784 Sofia, Bulgaria}
\altaffiltext{74}{Centre for Space Research, Private Bag X6001, North-West University, Potchefstroom Campus, Potchefstroom, 2520, South Africa}
\altaffiltext{75}{Graduate Institute of Astronomy, National Central University, 300 Zhongda Road, Zhongli 32001, Taiwan}
\altaffiltext{76}{Astronomical Observatory, Volgina 7, 11060 Belgrade, Serbia}
\altaffiltext{77}{Istanbul  University,  Science Faculty,  Department  of Astronomy and Space Sciences, Beyaz\'i t, 34119, Istanbul, Turkey}
\altaffiltext{78}{Aalto University Mets\"ahovi Radio Observatory, Mets\"ahovintie 114, FI-02540 Kylm\"al\"a, Finland}
\altaffiltext{79}{Institute of Astronomy and NAO, Bulgarian Academy of Sciences, 1784 Sofia, Bulgaria}
\altaffiltext{80}{Department of Physics, Brigham Young University Provo, UT}
\altaffiltext{81}{School of Cosmic Physics, Dublin Institute For Advanced Studies, Ireland}
\altaffiltext{82}{Institute for Astrophysical Research, Boston University, 725 Commonwealth Avenue, Boston, MA 02215}
\altaffiltext{83}{Astronomical Institute, St. Petersburg State University, Universitetskij Pr. 28, Petrodvorets,198504 St. Petersburg, Russia}
\altaffiltext{84}{Research Institute for Science and Engineering, Waseda University, 3-4-1, Okubo, Shinjuku, Tokyo 169-8555, Japan}
\altaffiltext{85}{Abastumani Observatory, Mt. Kanobili, 0301 Abastumani, Georgia}
\altaffiltext{86}{Engelhardt Astronomical Observatory, Kazan Federal University, Tatarstan, Russia}
\altaffiltext{87}{Aalto University Mets\"ahovi Radio Observatory, Mets\"ahovintie 114,  02540 Kylm\"al\"a, Finland}
\altaffiltext{88}{Aalto University Department of Radio Science and Engineering, P.O. BOX 13000, FI-00076 AALTO, Finland}
\altaffiltext{89}{ Institute of Astronomy with NAO, BAS, BG-1784, Sofia, Bulgaria}
\altaffiltext{90}{Astron.\ Inst., St.-Petersburg State Univ., Russia}
\altaffiltext{91}{Pulkovo Observatory, St.-Petersburg, Russia}
\altaffiltext{92}{Isaac Newton Institute of Chile, St.-Petersburg Branch}
\altaffiltext{93}{INAF, Osservatorio Astronomico di Torino, 10025 Pino Torinese (TO), Italy}
\altaffiltext{94}{Department of Physics, University of Colorado Denver Denver, CO}
\altaffiltext{95}{Astronomical Observatory, Volgina 7, 11060 Belgrade, Serbia}
\altaffiltext{96}{European Southern Observatory, Karl-Schwarzschild-Str. 2, 85748 Garching, Germany}

\begin{abstract}
We report on simultaneous broadband observations of the TeV-emitting blazar Markarian 501 between 1 April and 10 August 2013, including the first detailed characterization of the synchrotron peak with \textit{Swift} and \textit{NuSTAR}.  During the campaign, the nearby BL Lac object was observed in both a quiescent and an elevated state.  The broadband campaign includes observations with \textit{NuSTAR}, MAGIC, VERITAS, the \textit{Fermi} Large Area Telescope (LAT), \textit{Swift} X-ray Telescope and UV Optical Telescope, various ground-based optical instruments, including the GASP-WEBT program, as well as radio observations by OVRO, Mets\"ahovi and the F-Gamma consortium.  Some of the MAGIC observations were affected by a sand layer from the Saharan desert, and had to be corrected using event-by-event corrections derived with a LIDAR (LIght Detection And Ranging) facility. This is the first time that LIDAR information is used to produce a physics result with Cherenkov Telescope data taken during adverse atmospheric conditions, and hence sets a precedent for the current and future ground-based gamma-ray instruments. The \textit{NuSTAR} instrument provides unprecedented sensitivity in hard X-rays, showing the source to display a spectral energy distribution between 3 and 79 keV consistent with a log-parabolic spectrum and hard X-ray variability on hour timescales.  None (of the four extended \textit{NuSTAR} observations) shows evidence of the onset of inverse-Compton emission at hard X-ray energies.  We apply a single-zone equilibrium synchrotron self-Compton model to five simultaneous broadband spectral energy distributions.  We find that the synchrotron self-Compton model can reproduce the observed broadband states through a decrease in the magnetic field strength coinciding with an increase in the luminosity and hardness of the relativistic leptons responsible for the high-energy emission.   
\end{abstract}

\keywords{galaxies: BL Lacs --- galaxies: individual(Markarian 501) --- X-rays}

\section{Introduction}
Markarian\,501 (Mrk\,501) is a nearby, bright X-ray emitting blazar at $z=0.034$, also known to emit very-high-energy (VHE; $E\ge100$ GeV) gamma-ray photons \citep{quinn}.  Blazars are among the most extreme astrophysical sources, displaying highly variable emission at nearly every wavelength and timescale probed thus far.  These objects are understood to be active galactic nuclei that are powered by accretion onto supermassive black holes and have relativistic jets pointed along the Earth's line of sight \citep{urry}.  Relativistic charged particles within blazar jets are responsible for the non-thermal spectral energy distribution (SED) which is characterized by two broad peaks in the $\nu F_{\nu}$ spectral representation.  The origin of the lower-energy peak is relatively well understood, resulting from the synchrotron radiation of relativistic leptons in the presence of a tangled magnetic field \citep{marscherreview}.   Within the leptonic paradigm, the higher-energy SED peak is attributed to inverse-Compton up-scattering by the relativistic leptons within the jet of either the synchrotron photons themselves, namely synchrotron self-Compton (SSC) emission \citep{maraschi}, or a photon field external to the jet, namely external Compton (EC) emission \citep[e.g.][]{dermer, sikora}.  Alternatively, hadronic models attribute the higher-energy peak of blazar emission to proton synchrotron emission and/or synchrotron emission by secondary leptons produced in p-$\gamma$ interactions \citep{aharonian2002,bednarek}.

Along with the other nearby VHE blazar Mrk 421, Mrk\,501 represents one of the most comprehensively studied VHE blazars.  The blazar has been the subject of multiple broadband observation campaigns \citep[e.g.][]{catanese,kataoka,petry,abdoMrk501}.  Mrk\,501 is one of the brightest X-ray sources in the sky, and has been observed by \textit{RXTE} to display significant X-ray variability up to 20 keV \citep{gliozzi}.  During a phase of high activity at VHE energies in 1997, this source was also observed by \textit{BeppoSAX} to display unusually hard, correlated X-ray emission up to $>100$ keV, with a photon index of $\Gamma<2$ \citep{pian}. 

Observations of Mrk\,501 have so far lacked sufficient sensitivity at the hard X-ray energies (10-100 keV).  Observations at hard X-ray energies provide direct insight into the highest energy particles through detection of synchrotron emission.  There is also the possibility for insight into the lower energy particles through the detection of inverse-Compton emission from photon up-scattering by the lower-energy electrons.  As a relativistic synchrotron emitter, the falling edge of the synchrotron peak mimics the energy distribution of the emitting particles, allowing the highest energy particles to be directly probed through hard X-ray observations.  The energy-dependent cooling timescale can lead to more rapid variability at hard X-ray energies than at soft X-ray energies.  \cite{gliozzi} reported independent soft (2-10 keV) and hard (10-20 keV) X-ray variability of Mrk 501 using $RXTE$.  

Other hard X-ray observations have previously been performed with $BeppoSAX$ \citep{massaro2004a} and $Suzaku$ HXD \citep{mrk501MAGIC2}.  Due to the rapid X-ray variability displayed by blazars such as Mrk 501, the long integration time required for significant detection and spectral reconstruction by the aforementioned X-ray instruments was not ideal for extracting information about hard X-ray variability.  Much more sensitive hard X-ray observations of blazars, however, are now possible with Nuclear Spectroscopic Telescope Array \textit{NuSTAR}.

\textit{NuSTAR} is a hard X-ray (3-79 keV) observatory launched into a low Earth orbit in June 2012 \citep{fiona}.  It features the first focusing hard X-ray telescope in orbit that allows high sensitivity beyond the 10 keV cutoff shared by all other currently active focusing soft X-ray telescopes. The inherently low background associated with concentrating the X-ray light enables \textit{NuSTAR} to achieve approximately a one-hundred-fold improvement in sensitivity over the collimated and coded-mask instruments that operate in the same spectral range. 

\textit{NuSTAR} observed Mrk\,501 four times in 2013 as part of a simultaneous multiwavelength (MWL) campaign, including VHE observations by MAGIC and VERITAS, high-energy (HE; 100 MeV-100 GeV) gamma-ray observations by the \textit{Fermi} Large Area Telescope (LAT), soft X-ray and UV observations with \textit{Swift} X-ray Telescope (XRT) and Ultraviolet Optical Telescope (UVOT), optical observations from a number of ground-based instruments including the GASP-WEBT program, as well as radio observations by the Owens Valley Radio Observatory (OVRO; 15 GHz), Mets\"ahovi (37 GHz) and the F-Gamma monitoring program, providing measurements between 2.64 GHz and 228.39 GHz.  The \textit{NuSTAR} observations took place on 2013 April 13, 2013 May 8, 2013 July 12 and 13 (MJD 56395, 56420, 56485 and 56486, respectively), with the latter two observations resulting from target of opportunity (ToO) exposures triggered by an elevated state observed by the \textit{Swift} XRT and the MAGIC telescopes.  

We use these observations to study the hard X-ray spectral behavior of Mrk 501 in detail over multiple flux states.  The \textit{NuSTAR} observations, analysis and results are detailed in Section 2, with the contemporaneous MWL observations, analysis and results shared in Section 3.  After comparing the simultaneous \textit{Swift} XRT and \textit{NuSTAR} observations in Section 4, we investigate variability of the source in Section 5.  The MWL SEDs are constructed over the multiple observed states and investigated in terms of a single-zone equilibrium synchrotron self-Compton model in Section 6, with discussion and conclusions provided in Section 7.

\section{\textit{NuSTAR} Observations and Analysis}
In order to maximize the strictly simultaneous overlap of observations by \textit{NuSTAR} and ground-based VHE observatories during this broadband campaign of Mrk 501, the observations were arranged according to visibility of the blazar at the MAGIC and VERITAS sites. The \textit{NuSTAR}  coordinated observations involving both VERITAS and MAGIC were performed on 2013 April 13 and 2013 May 8, with the \textit{NuSTAR} ToO observations (initiated by \textit{Swift} and MAGIC) performed on 2013 July 12 and 13.   The \textit{NuSTAR} observations typically spanned 10 hours, resulting in 10-30 ks of source exposure after removing periods of orbital non-visibility.  The observation details are summarized in Table~1.  The data were reduced using the standard \texttt{NuSTARDAS} software package\footnote{\texttt{http://heasarc.gsfc.nasa.gov/docs/nustar/analysis/}} \texttt{v1.3.1}.

\begin{deluxetable*}{ccccc}
\tablecolumns{5}
\tablewidth{0pc}
\scriptsize
\tablecaption{Summary of the \textit{NuSTAR} hard X-ray observations of Mrk 501.  The observations are sometimes referred to with the last three digits of the Observation ID within this work. }
\scriptsize
\tablehead{
\colhead{Observation} & \colhead{MJD Exposure}       & \colhead{Exposure} & \colhead{Number} & \colhead{Detection} \\
\colhead{ID} & \colhead{Range}     & \colhead{[ks]} & \colhead{Orbits} & \colhead{Range [keV]} }
\startdata
60002024002&56395.1-56395.5&19.7&6&3-60\\
60002024004&56420.8-56421.5&28.3&10&3-65 \\
60002024006&56485.9-56486.2&11.9&4&3-70 \\
60002024008&56486.8-56487.1&11.4&4&3-70 \\
\enddata
\end{deluxetable*}

\begin{scriptsize}
\begin{deluxetable*}{ccc|cccc|cccccc}
\setlength{\tabcolsep}{-0.2in} 
\tabletypesize{\scriptsize}
\tablecolumns{10}
\tablewidth{0pc}
\scriptsize
\tablecaption{\textit{NuSTAR} spectral fit summary, with integral flux values (in units of $\times 10^{-11}$ erg cm$^{-2}$ s$^{-1}$) derived from the log-parabolic fits.  Data, models and ratios are shown in Figure 1.  The indices of the LP fits are derived at 10 keV.  The errors for each parameter are found using a value of $\Delta\chi^2$=2.706, corresponding to a 90\% confidence level for a parameter. Observation IDs are shortened by removing the first 60002024 identifier in column one. }
\scriptsize
\tablehead{
\colhead{}    &  \multicolumn{2}{c|}{Power law} &  
\multicolumn{4}{c|}{Broken Power law} & \multicolumn{5}{c}{Log Parabola} \\
\cline{2-3} \cline{4-7} \cline{8-12} \\
\colhead{Obs.} & \colhead{Index}       & \colhead{PL} &  \colhead{Index} & \colhead{Index} & \colhead{$E_{\rm break}$}& \colhead{BKNPL}
     & \colhead{Index}   & \colhead{Curvature}& \colhead{LP} & \colhead{3-7 keV} & \colhead{7-30 keV}\\
\colhead{ID} & \colhead{$\Gamma$}     & \colhead{$\chi^2$/DOF}  & \colhead{$\Gamma_1$} & \colhead{$\Gamma_2$}&\colhead{[keV]} & \colhead{$\chi^2$/DOF} &
     \colhead{$\Gamma$}   & \colhead{$\beta$} & \colhead{$\chi^2$/DOF} & \colhead{Flux} & \colhead{Flux}}
\startdata
002 &2.216$\pm$0.009 &831/700&2.04$\pm$0.03&2.34$\pm$0.02&6.3$\pm$0.4&747/698&2.290$\pm$0.010&0.26$\pm$0.03&729/699&3.72$\pm$0.02&4.81$\pm0.03$\\
004& 2.191$\pm$0.006&1204/889&1.25$\pm$0.20&2.21$\pm$0.01&3.1$\pm$0.1&1211/887&2.250$\pm$0.008&0.21$\pm$0.02&1051/888&5.19$\pm$0.02&6.98$\pm$0.05\\
006 &2.060$\pm$0.006 &1246/924&1.92$\pm$0.02&2.22$\pm$0.02&7.9$\pm$0.4&1057/922 &2.115$\pm$0.008&0.24$\pm$0.02&1024/923&12.08$\pm$0.09&18.6$\pm$0.1\\
008 & 2.081$\pm$0.007&1152/863&1.90$\pm$0.02&2.25$\pm$0.02&7.4$\pm$0.3&914/861&2.149$\pm$0.008&0.32$\pm$0.02&892/862&10.75$\pm$0.05&16.4$\pm$0.1\\
\enddata
\end{deluxetable*}
\end{scriptsize}

\begin{figure}
\epsscale{1.3}
\center\includegraphics[width=3in]{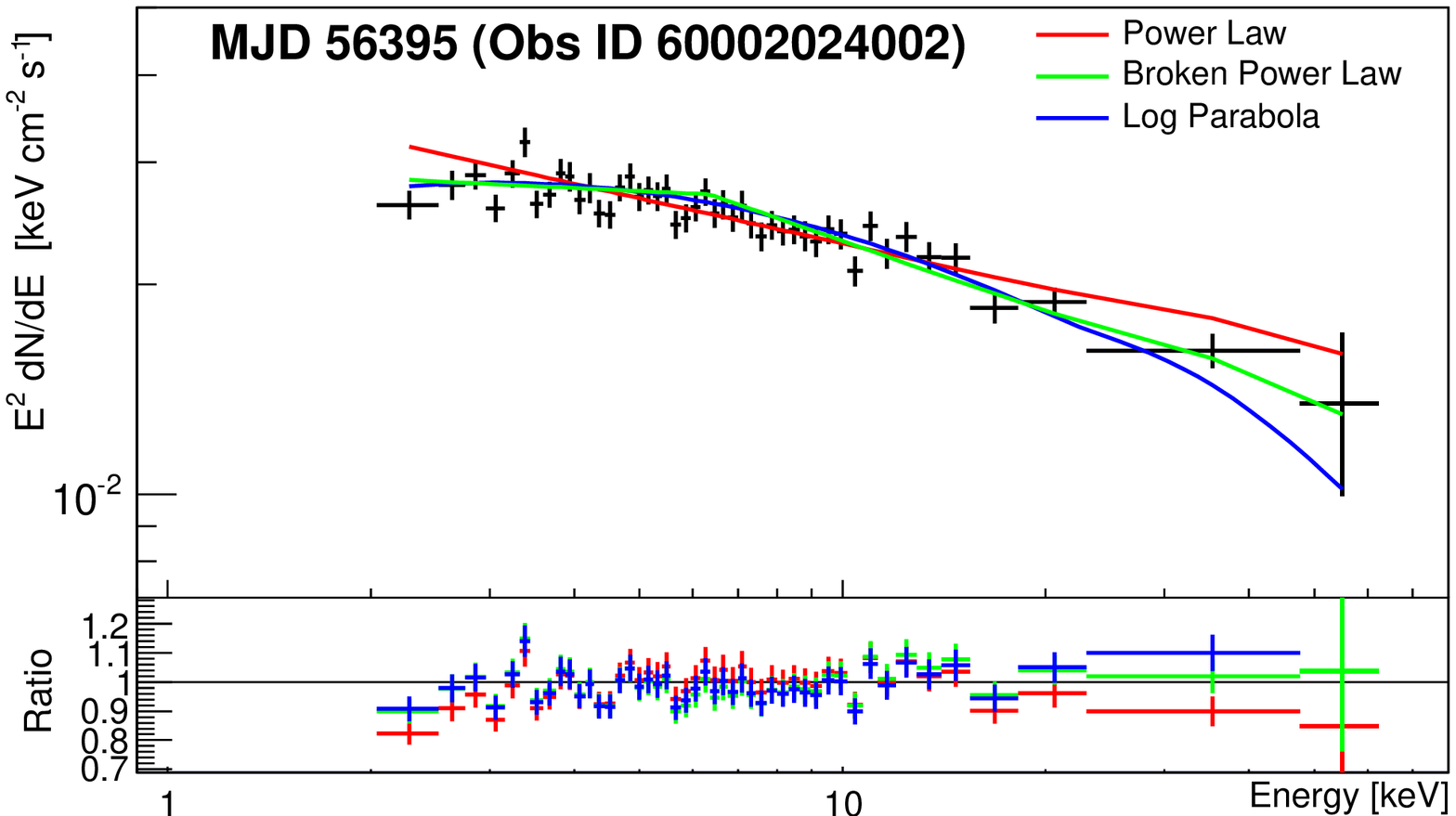}
\includegraphics[width=3in]{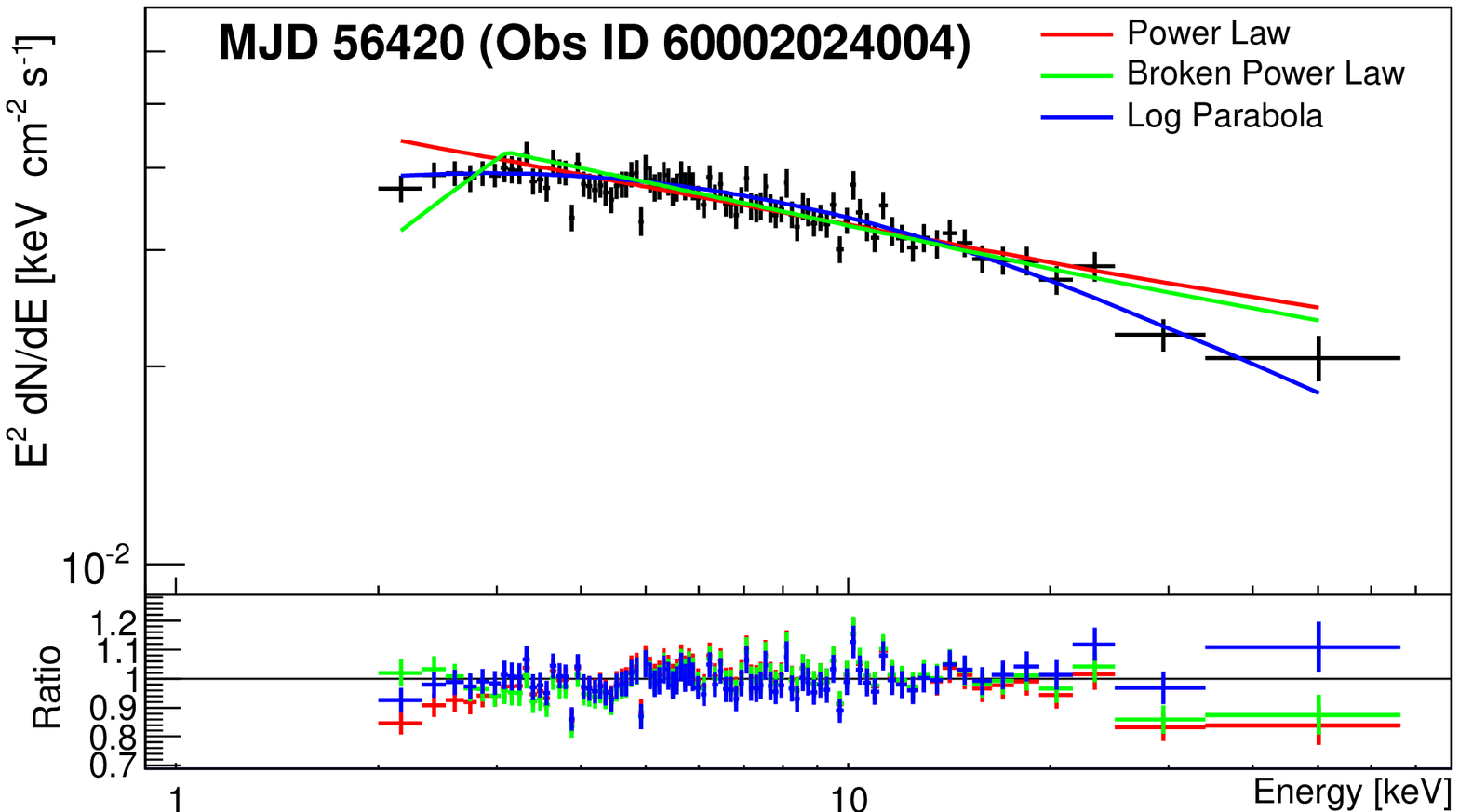}\\
\includegraphics[width=3in]{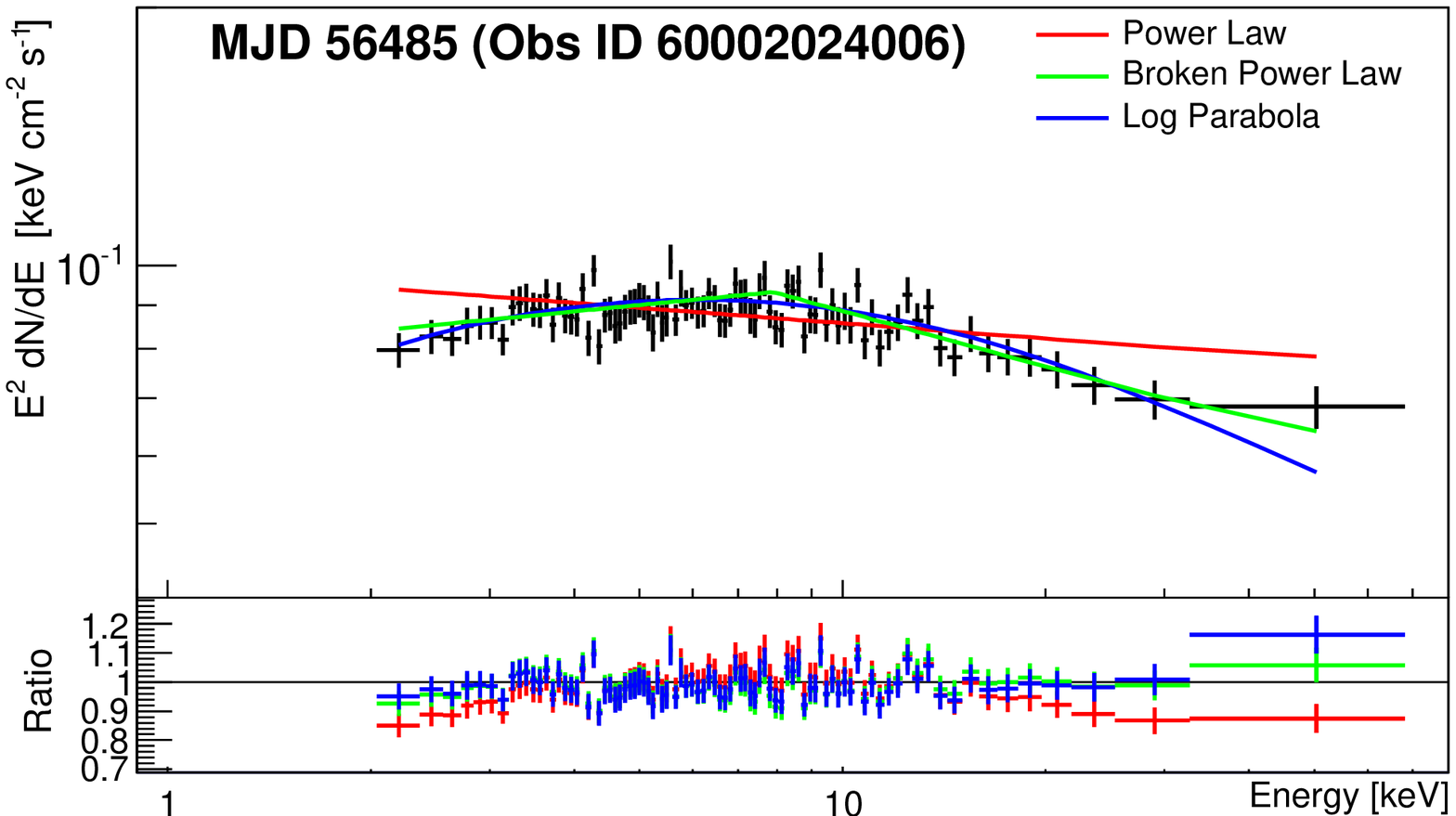}
\includegraphics[width=3in]{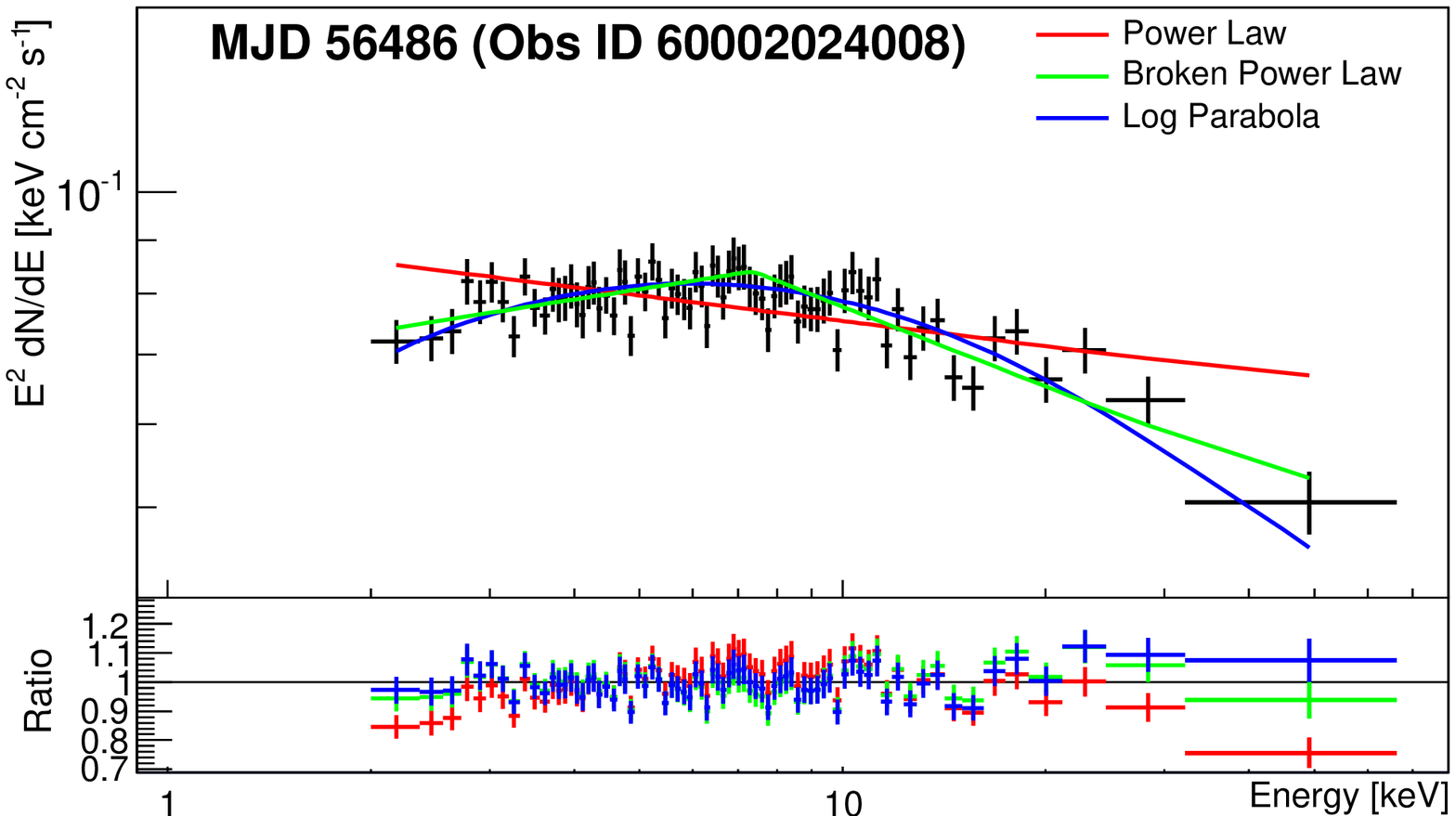}\\
\caption{The spectral energy distributions of Mrk 501 derived from the \textit{Nu-STAR} observations, showing the PL (red), BKNPL (green) and LP (blue) models fitted to each observation.  The \textit{NuSTAR} observations show significant detection of the blazar up to at least 65 keV in each observation.  The data-to-model ratios are shown in the bottom panel of each plot, with the spectral fit parameters summarized in Table 2. Spectra have been rebinned for figure clarity.}\label{fig1}
\end{figure}

The spectral analysis was performed with \texttt{XSPEC} \footnote{\tt http://heasarc.nasa.gov/docs/software/lheasoft/xanadu/\\xspec/XspecManual.pdf} 
Version 12.7.1.  The data were binned to require 20 counts per bin, and fit with three spectral models via $\chi^2$ minimization.   The first model applied to the data is a power law 

\begin{equation}
 A(E)_{\rm PL}= K (E/E_\mathrm{0})^{-\Gamma} ,
 \end{equation}
 
\noindent referred to as the PL model for the remainder of this work, where $F(E)$ is the flux at energy $E$, $\Gamma$ is the index, $K$ is the normalization parameter (in units of photons keV$^{-1}$cm$^{-2}$s$^{-1}$) and $E_0$ is fixed at 10 keV.

The second spectral model applied to the data is a broken power law, referred to as BKNPL model for the remainder of this work.  The model is made up of two power-law photon indices, meeting at a break energy $E_\mathrm{break}$
\begin{equation}
A(E)_{\rm BKNPL}= K (E/E_\mathrm{break})^{-\Gamma_{1,2}}
\end{equation}

\noindent where $\Gamma_{1}$ and $\Gamma_{2}$ represent the photon indices below and above the break energy $E_\mathrm{break}$, respectively.

The third spectral model applied to the data is a log parabola, referred to as the LP model for the remainder of this work. This model has been suggested to better represent the X-ray spectra of TeV-detected blazars between 0.2 and 100 keV \citep[e.g.][]{massaro2004b, tramacere2007a}.   This model allows the spectral index to vary as a function of energy according to the expression 

\begin{equation}
A(E)_{\rm LP}=K(E/E_\mathrm{0})^{-(\Gamma+\beta {\rm log}(E/E_0))},
 \end{equation}

\noindent with a curvature parameter $\beta$.  The spectral data, model fits and data-to-model ratios for each \textit{NuSTAR} observation are shown in Figure~1.  The spectral fitting results for each model as applied to the \textit{NuSTAR} observations are summarized in Table~2.  The errors for each parameter are found using a value of $\Delta\chi^2$=2.706, corresponding to a 90\% confidence level for one parameter. 

For all four \textit{NuSTAR} observations, the X-ray emission of Mrk 501 is best represented with a log parabola.  A statistical $F-$test \citep{snedecor} using the $\chi^2$ and degrees of freedom (DOF) of the PL versus LP fit results in $F$-statistics of 97.8, 129.3, 200.1 and 251.3 for the observations 002, 004, 006 and 008, respectively, corresponding to probabilities of 1.1$\times10^{-21}$, 4.6$\times10^{-28}$, 2.9$\times10^{-41}$ and 7.9$\times10^{-50}$ for being consistent with the null PL hypothesis.   The broken power-law fit to the second \textit{NuSTAR} observation, ID 004, produces a break energy at the lower limit of the \textit{NuSTAR} sensitivity window, and is interpreted as a failed fit.    The other three observations fit the break energy near $E_{\rm break}$=7 keV, motivating the decision to present the \textit{NuSTAR} flux values in the 3-7 keV and 7-30 keV bands throughout this work.  The upper bound of 30 keV is the typical orbit-timescale detection limit for the Mrk 501 observations.

The \textit{NuSTAR} observations show the blazar to be in a relatively low state for the first two observations, and a relatively high state during the last two observations, with the 3-7 keV integral fluxes derived from the log-parabolic fits 2-4 times higher than found for the first two observations.  More specifically, the average 3-7 keV integral flux values (in units of $10^{-11}$ erg cm$^{-2}$ s$^{-1}$) were 3.72$\pm$0.02 and 5.19$\pm$0.02, respectively, for the observations occurring on MJD 56395 and 56420, and 12.08$\pm$0.09 and 10.75$\pm$0.05, respectively, for the observations starting on MJD 56485 and 56486.  In the same flux units, the 7-30 keV integral flux values for the first two observations are similarly 3-4 times lower than the flux states observed in the last two observations (4.81$\pm$0.03 and 6.98$\pm$0.05 on MJD 56395 and 56420 as compared to 18.6$\pm$0.1  and 16.4$\pm$0.1 on MJD 56485 and 56486).  These integral flux values are summarized in Table~2.

The \textit{NuSTAR} observations extend across multiple occultations by the Earth, and the integral flux and index ($\Gamma$) light curves for the orbits of each extended observation are shown in Figure~2.  The periods with simultaneous observations with the ground-based TeV instruments of MAGIC and VERITAS are highlighted by grey and brown bands in the upper portion of each light curve.  The observations and results from MAGIC and VERITAS for these time periods are summarized in Section 3.1.

\begin{figure}
\epsscale{0.8}
\center\includegraphics[width=3.1in]{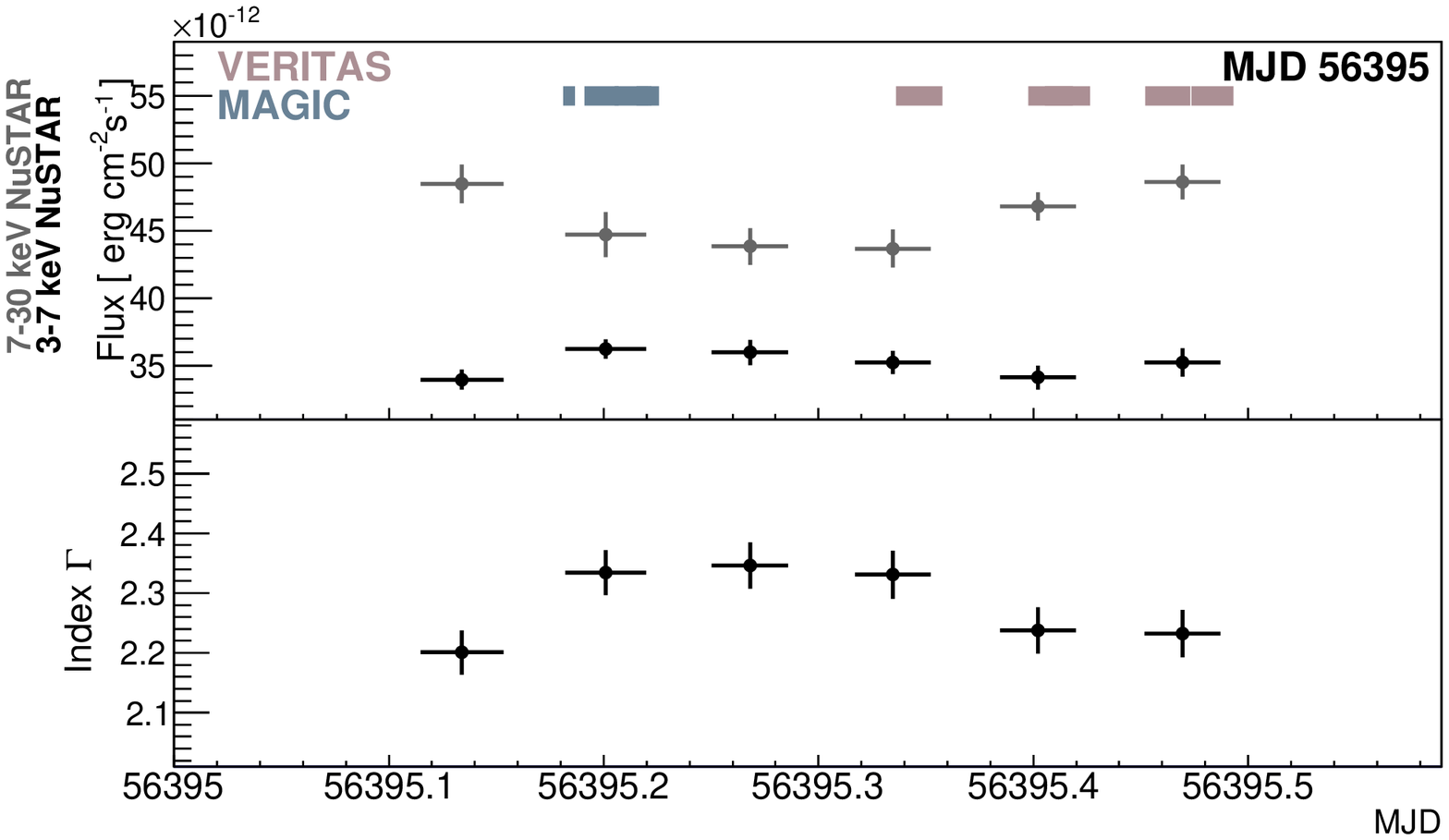}\\
\includegraphics[width=3.1in]{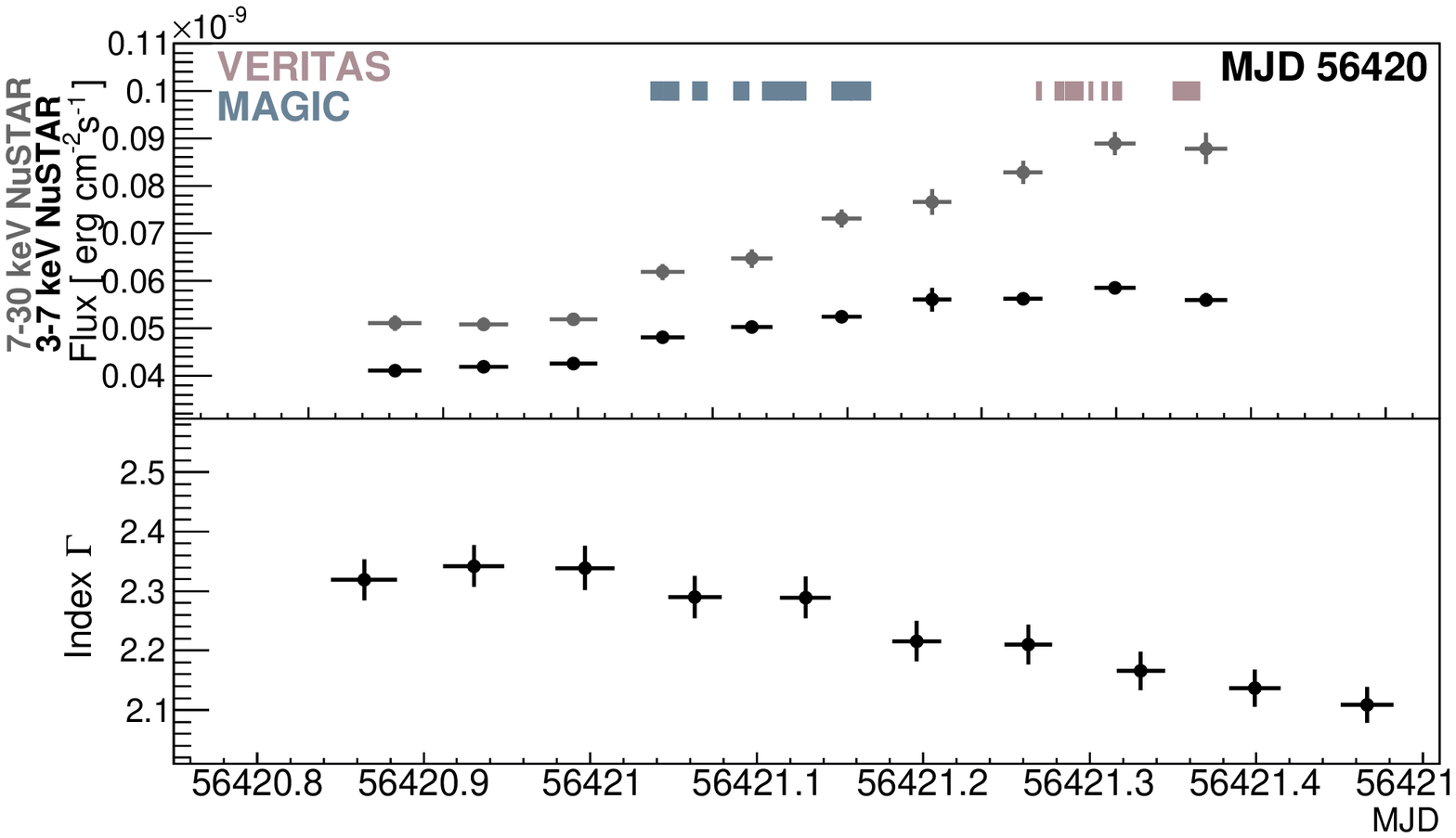}\\
\includegraphics[width=3.1in]{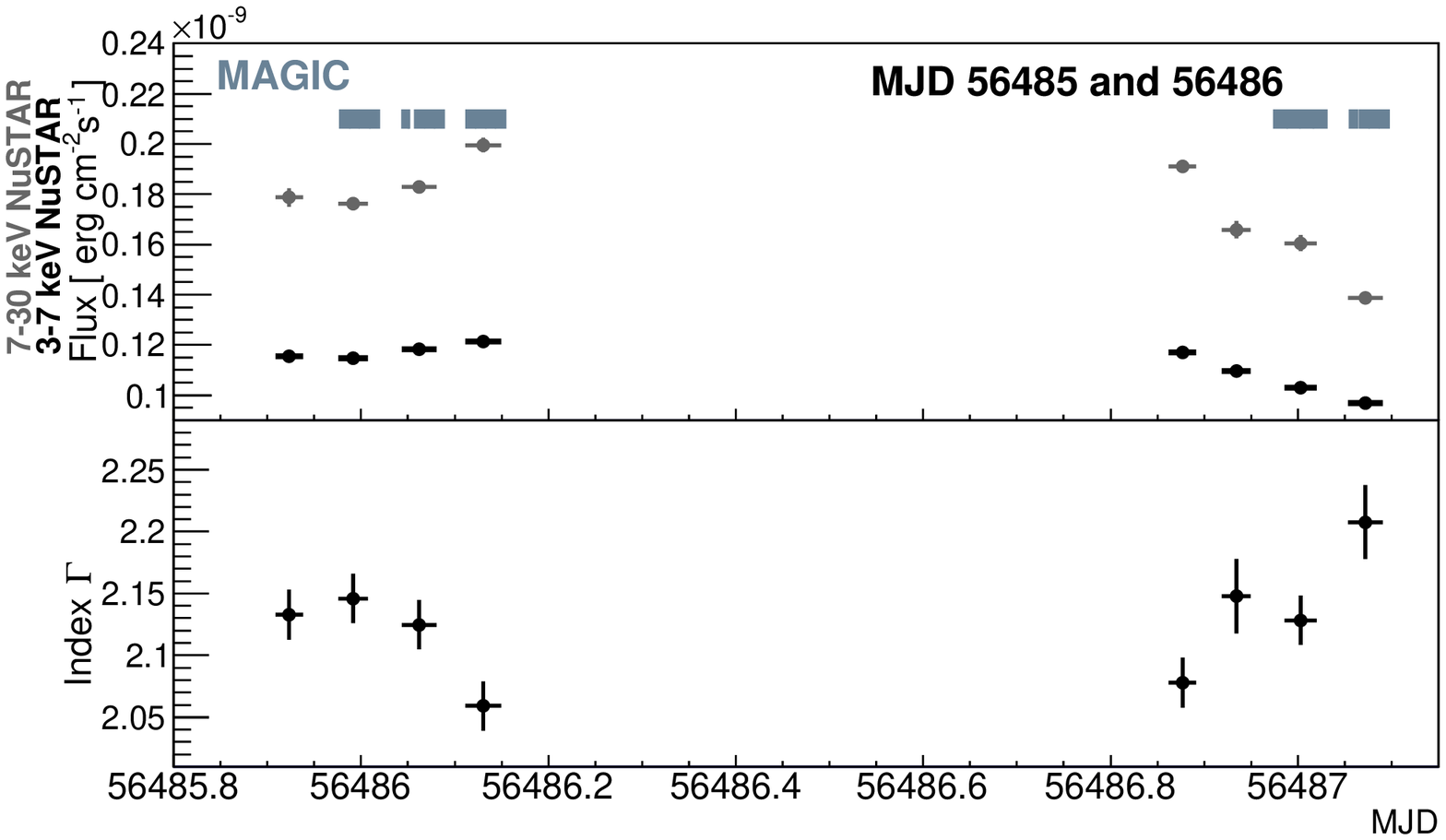}\\
\caption{The \textit{NuSTAR} orbit-binned light curves, with 3-7 keV (black) and 7-30 keV (grey) integral flux values (top panel of each plot) and the log-parabolic indices ($\Gamma$, lower panel) with the curvature parameters ($\beta$) fixed to the value found for the full \textit{NuSTAR} exposure. The third and fourth observations are shown in the third plot.  The periods where simultaneous quality-selected observations with MAGIC and VERITAS occurred are highlighted in the top panel of each plot with color coded bands. We note that the vertical axes are set differently for each observation to allow a clear view of the orbit-to-orbit variability and that the light curve for the full campaign is shown in Figure 5.}\label{fig1}
\end{figure}

The 3-7 keV and 7-30 keV integral flux values of the first exposure (Observation ID 002) show low variability ($\chi^2=7.0$ and 13.4 for 5 DOF), while the trend of increasing flux in both the 3-7 keV and 7-30 keV bands is clear during the second observation (Observation ID 004).  The 7-30 keV flux increases from (5.1$\pm$0.1) $\times10^{-11}$ erg cm$^{-2}$ s$^{-1}$ to (8.8$\pm$0.1) $\times10^{-11}$ erg cm$^{-2}$ s$^{-1}$ in fewer than 16 hours.  The 7-30 keV increases from (1.7$\pm$0.1) $\times10^{-10}$ erg cm$^{-2}$ s$^{-1}$ to (2.0$\pm$0.1) $\times10^{-10}$ erg cm$^{-2}$ s$^{-1}$ in fewer than 7 hours on MJD 56485 (Observation ID 006) and significantly decreases from (1.9$\pm$0.1) $\times10^{-10}$ erg cm$^{-2}$ s$^{-1}$ to (1.4$\pm$0.1) $\times10^{-10}$ erg cm$^{-2}$ s$^{-1}$, again in fewer than 7 hours on MJD 56486 (Observation ID 008).

\begin{figure}
\epsscale{1.3}
\center\includegraphics[width=3.2in]{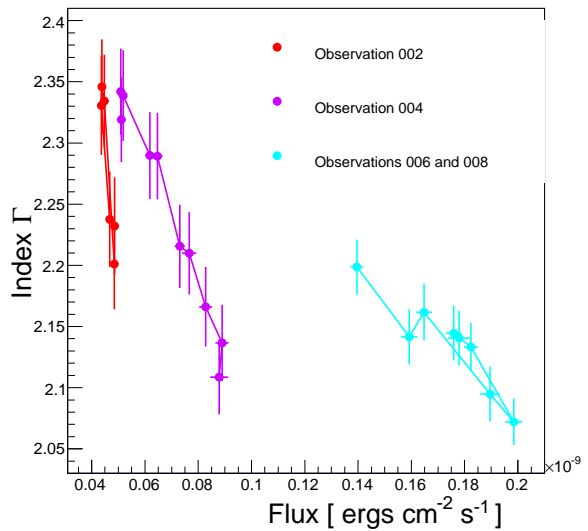}\\
\caption{The log-parabolic fit index $\Gamma$ at 10 keV versus the 7-30 keV integral flux for \textit{NuSTAR}, binned by orbit.  The first exposure is shown in red, the second in violet, and the last two in cyan, with solid lines meant to guide the eye along the parameter evolution over the full observations.  In all three cases, the spectrum hardens when the intensity increases; in the fourth observation, the spectrum then softens as the intensity decreases. }\label{fig1}
\end{figure}

The relation between the log-parabolic photon indices and 7-30 keV flux values resulting from the fits to the \textit{NuSTAR} observations of Mrk 501 are shown for each observation separately in Figure 3.  
The curvature $\beta$ was not seen to change significantly from orbit to orbit and therefore was fixed at the average value found for each observation (see Table~2 for values).  The count rate light curves show no indications of variability on a timescale of less than an orbit period ($\sim$90 minutes).  As observed previously in the X-ray band for Mrk 501 \citep{kataoka}, the source was displaying a harder-when-brighter trend during this campaign. This has also been observed in the past for Mrk 421 \citep{hardwhenbright}.

\section{Broadband Observations}
\subsection{Very-High-Energy Gamma Rays}
\subsubsection{MAGIC}

MAGIC is a VHE instrument composed of two imaging atmospheric Cherenkov telescopes (IACTs) with mirror diameters of 17 m, located at 2200 m above sea level at the Roque de Los Muchachos Observatory on La Palma, Canary Islands, Spain. The energy threshold of the system is 50 GeV and it reaches an integral sensitivity of 0.66\% of the Crab Nebula flux above 220 GeV with a 50-hour observation \citep{aleksicUpgrade}.

MAGIC observed Mrk 501 in 2013 from April 9 (MJD 56391) to August 10 (MJD 56514). On July 11 (MJD 56484), ToO observations were triggered by the high count rate of $\sim$15 counts s$^{-1}$ observed by \textit{Swift} XRT (see Section 3.3).  The flaring state was observed intensively for five consecutive nights until July 15 (MJD 56488). After that the observations continued with a lower cadence until August 10. 

The source was observed during 17 nights, collecting a total of 22 hours of data with zenith angles between 10$^{\circ}$ and 60$^{\circ}$. Only five hours survived the standard quality cuts for regular MAGIC data analysis because many observations were taken during the presence of a Saharan sand-dust layer in the atmosphere known as ``Calima".  As we explain below, using the LIDAR information we could recover 10 of the 17 hours which would have been rejected otherwise. The telescopes were operated in the so-called wobble mode \citep{fomin}, where the pointing direction is changed every 20 (or 15) minutes among 2 (or 4) positions with an offset of $0.4^{\circ}$ from the source position. 

\begin{figure}
\epsscale{1.3}
\center\includegraphics[width=3.2in]{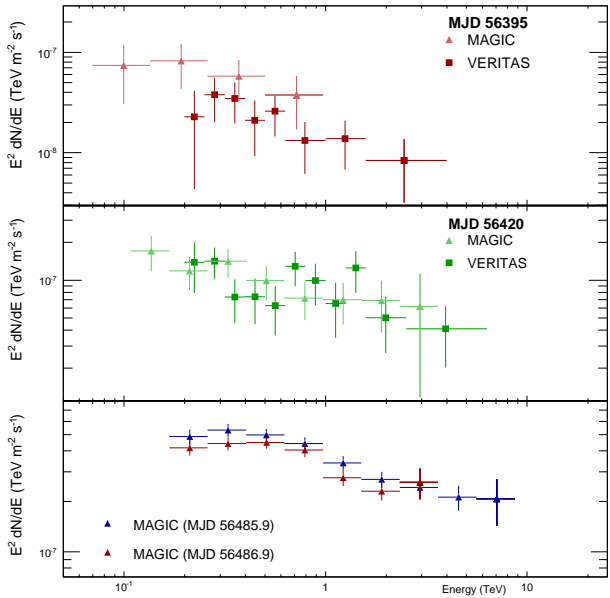}
\caption{MAGIC and VERITAS spectra averaged over epochs with simultaneous \textit{NuSTAR} exposures.  The power-law spectral fitting parameters for the VHE data are summarized in Table~3.  Only statistical (1$\sigma$) error bars are shown for each of the spectral points.} 
\end{figure}

\begin{scriptsize}
\begin{deluxetable*}{cccccccccc}
\tabletypesize{\scriptsize}
\tablecolumns{12}
\tablewidth{0pc}
\scriptsize
\tablecaption{MAGIC and VERITAS observations, analysis and spectral fit summary for \textit{NuSTAR}-simultaneous observations.  Observations occurring on the same day are grouped with horizontal lines.  Daily average values of MAGIC observations are shown in bold, below the results for each observation occurring on that day.  Statistical (1$\sigma$) error bars are provided for the power-law indices and the integral fluxes. The flux value between MJD 56486.106 and 56486.148 (shown in italics) is estimated with fitting parameters due to an energy threshold above 200 GeV.  The significance of the observed gamma-ray signals is computed according to Eqn. 17 in \cite{lima}. }
\scriptsize
\tablehead{
\colhead{Exposure} & \colhead{Exposure}       & \colhead{Exposure} & \colhead{Instrument} & \colhead{Zenith} & \colhead{Detection} &\colhead{Power-law}       & \colhead{Integral Flux}   & \colhead{$\chi^2$}& \colhead{DOF}\\
\colhead{Start MJD} & \colhead{Stop MJD}     & \colhead{Length} & \colhead{}  & \colhead{angle} & \colhead{Significance} & \colhead{Index}        & \colhead{$>$ 200 GeV}   & \colhead{} & \colhead{}\\
\colhead{} & \colhead{}     & \colhead{[hr]} & \colhead{} & \colhead{[deg]} & \colhead{[$\sigma$]} & \colhead{}        & \colhead{ [$\times10^{-11}$ ph cm$^{-2}$s$^{-1}$]}   & \colhead{} & \colhead{}
}
\startdata
56395.179	&56395.223	&1.0&	MAGIC	& 10-14 &	7.8&	2.50$\pm$0.24                        &2.39$\pm$0.44	  &   0.58      & 6\\
56395.336		&56395.493		&2.5&				     VERITAS&	15-35 &	  8.3&	     3.1$\pm$0.4 &1.85$\pm$0.38    &0.76 & 5\\
\hline
56421.142 &   56421.209 & 1.1&  MAGIC     & 12-28 & 12.5    &   2.24$\pm$0.08             & 5.08$\pm$0.54  & 15.5  & 13\\
56421.340 &56421.462&1.0& VERITAS&	  20-32 & 14.7&	   2.25$\pm$0.15&4.45$\pm$0.61   &		  6.9&	  9\\
\hline
56485.972	&56486.014&1.0		&         MAGIC	     &12-24 &20.4	        &    2.19$\pm$0.07                    &20.8$\pm$1.2   & 10.0        &12\\
56486.039	&56486.083&		1.0	  &         MAGIC  &	28-43 & 20.7       &			        2.39$\pm$0.08        &25.2$\pm$1.3   &     26.5    & 10 \\
56486.106        &56486.148&  1.0   &        MAGIC &  48-60 &  14.3              &  2.71$\pm$0.12          & \textit{32.4$\pm$2.0}   & 11.9  &11 \\
\textbf{56485.972}& \textbf{56486.148} & 2.9& MAGIC & 12-60 & 32.3 & 2.28$\pm$0.04& 24.3$\pm$0.8 &24.1 &15 \\
\hline
56486.966	&56487.022&	1.3	&         MAGIC& 12-27 &25.2          &      2.37$\pm$0.06                   &24.9$\pm$1.1   &  20.3       & 12\\
56487.050	&56487.091&	0.9	&         MAGIC& 33-46 &    18.5       &     2.23$\pm$0.09                   & 17.8$\pm$1.0	  &  14.5       & 11\\
\textbf{56486.966}		&\textbf{56487.091}  & 2.2   &   MAGIC & 12-46 &31.8    &         2.31$\pm$0.05               &		  20.9$\pm$0.7 & 30.4        & 12\\
%
\enddata
\end{deluxetable*}
\end{scriptsize}

All the data were analyzed following the standard procedure \citep{aleksic2012} using the MAGIC Analysis and Reconstruction Software (MARS; \citealt{zanin}). An image cleaning was applied based on information of signal amplitude and timing of each pixel, and the shower images were parametrized using the Hillas parameters \citep{hillas}.  For the reconstruction of the gamma-ray direction and the gamma-hadron separation, the random forest method is applied using the image parameters and the stereoscopic parameters. \citep{albert2008, aleksic2010}. The energy reconstruction utilizes look-up tables. The analysis steps were confirmed independently with data from the Crab Nebula and dedicated Monte Carlo simulations of gamma-ray showers.   

A fraction of the dataset (10.4 of 15.1 hours, specifically the observations between MJD 56485 and MJD 56514) was affected by ``Calima,'' a Saharan sand-dust layer in the atmosphere. A correction within the framework of the MARS software is applied to account for the absorption due to Calima using LIDAR measurements taken simultaneously with the MAGIC observations \citep{fruck}. 
The correction was carried out in two steps. Due to the dust attenuation during Calima, the estimated energy is shifted towards low energies, and thus is corrected event by event, as the first step.  Then, to account for the shift of the energy estimation, a correction to the collection area is applied as a second step, due to the energy dependence in the collection area. The atmospheric transmission values for this method were obtained from the temporally closest LIDAR measurement. During the observations affected by Calima the atmospheric transmission ranged from 85\% down to 60\%, being relatively stable within a timescale of one day, which is a typical feature of a Calima layer (unlike a cloudy sky). The precision on the energy correction is estimated to be around 5\% of the attenuation (40\% to 15\%), which corresponds to $<2\%$ of the estimated energy, at most. After the Calima correction, the energy threshold increases inversely proportional to the transmission value. This correction method was tested independently on a Crab Nebula dataset observed under similarly hazy weather conditions \citep{ATOM}. Details of the method can be found in \cite{fruckthesis}. This is the first time an event-by-event atmospheric correction is applied to MAGIC data.

The analysis results of the MAGIC data taken during good weather conditions have a systematic uncertainty in the  flux normalization and in the energy scale. For both of them, the component changing run-by-run is estimated to be $\sim$11\% using Crab Nebula observations \citep{aleksicUpgrade}. 
It is attributed mainly to the atmospheric transmission of the Cherenkov light, which can change on a daily basis (even during so-called good weather conditions) and the mirror reflectivity, which can change also on a daily basis due to the deposition of dust.  
The atmospheric correction applied in the analysis of the data taken during Calima increases this run-by-run systematic error from 11\% to 15\% due to the uncertainty in the correction.  
Since the systematic uncertainty can be different according to the atmospheric correction,we have added 15\% or 11\% (with or without the atmospheric correction) to the statistical errors of the flux in quadrature for the evaluation of flux variability. 

The summary of the MAGIC analysis results for observations occurring simultaneously with \textit{NuSTAR} is provided in Table 3. 
The derived spectra are shown in Figure 4, where the spectral points are drawn with statistical errors only. 
The resultant flux values above 200 GeV range from $(2.39 \pm  0.51) \times 10^{-11}$  ph cm$^{-2}$ s$^{-1}$ (0.11 Crab Nebula flux) on MJD 56395 to $(5.52 \pm  0.87) \times 10^{-10}$  ph cm$^{-2}$ s$^{-1}$ (2.5 times the Crab Nebula flux) on MJD 56484.  
As seen in the overall light curve (top panel of Fig. 5, shown again only with statistical errors), MAGIC observations indicate a significant variability around MJD 56484.  
A hint of intra-night variability was observed on MJD 56486 and 56487 simultaneously with the \textit{NuSTAR} observations, as shown in the zoomed-in light curve (top panel of Figure 6). During these two nights the VHE emission is consistent with a constant flux, resulting in a $\chi^2$/DOF of 7.3/4 (12\% probability) with the inclusion of the systematic error.  Without accounting for the additional systematic error, the constant fit to the flux results in a $\chi^2$/DOF of 57/4.

\subsubsection{VERITAS}

VERITAS is a VHE instrument comprised of four 12-m IACTs and is sensitive to gamma rays between $\sim$100 GeV and $\sim$30 TeV \citep{holder2006, keida}.  This instrument can detect 1\% Crab Nebula flux in under 25 hours.  VERITAS observed Mrk 501 fourteen times between 2013 April 7 (MJD 56389) and 2013 June 18 (MJD 56461), with 2.5 and 1.0 hours quality-selected exposures occurring simultaneously with \textit{NuSTAR} on MJD 56395 and MJD 56421, respectively.   On days without simultaneous \textit{NuSTAR} observations, the exposure times ranged between 0.5 hours and 1.5 hours.  The observations occurring simultaneously with \textit{NuSTAR} are summarized in Table 3.  Due to an annual, $\sim$2 month long monsoon season in southern Arizona where VERITAS is located, no VERITAS observations were possible for this campaign after 2013 June 18.  

The VERITAS observations were taken with 0.5$^{\circ}$ offset in each of the four cardinal directions to enable simultaneous background estimation \citep{fomin}.   Events were reconstructed following the procedure outlined in \cite{acciari2008}.  The recorded shower images were parameterized by their principal moments, giving an efficient suppression of the far more abundant cosmic-ray background.  Cuts were applied to the mean scaled width, mean scaled length, apparent altitude of the maximum Cherenkov emission (shower maximum), and $\theta$, the angular distance between the position of Mrk 501 and the reconstructed origin of the event.   The results were independently reproduced with two analysis packages \citep{cogan,heike}.  The uncertainty on the energy calibration of VERITAS is estimated at 20\%.  Additionally, the systematic uncertainty on the spectral index is estimated at 0.2, appearing to be relatively independent of the source slope \citep{madhavan}.

A differential power law is fit to the data ($dN/dE\propto E^{-\Gamma}$) to characterize the VHE spectrum of the source.  VERITAS observed Mrk 501 to vary by no more than a factor of three in flux throughout the observations, with the integral flux ranging from (1.85$\pm$0.38$)\times\,10^{-11}$ ph cm$^{-2}$s$^{-1}$  above 200 GeV (8\% Crab Nebula flux above the same threshold) on MJD 56395 to (4.45$\pm$0.61$)\times\,10^{-11}$ ph cm$^{-2}$s$^{-1}$ (20\% Crab Nebula flux) on MJD 56421.  The source displayed low spectral variability, ranging between $\Gamma=3.1\pm0.4$ in the low flux state to $\Gamma=2.19\pm0.07$ in the higher flux state.  The observation and analysis results are summarized in Table 3 (for \textit{NuSTAR} simultaneous observations only), with the VHE spectra of the \textit{NuSTAR} simultaneous observations shown in Figure 4. Day-to-day uncertainties in flux calculations that might be introduced by different atmospheric conditions (even under strictly good weather conditions) are not included in Table 3 and are estimated at less than 10\%.

\begin{figure*}
\epsscale{0.9}
\center \includegraphics[width=5in]{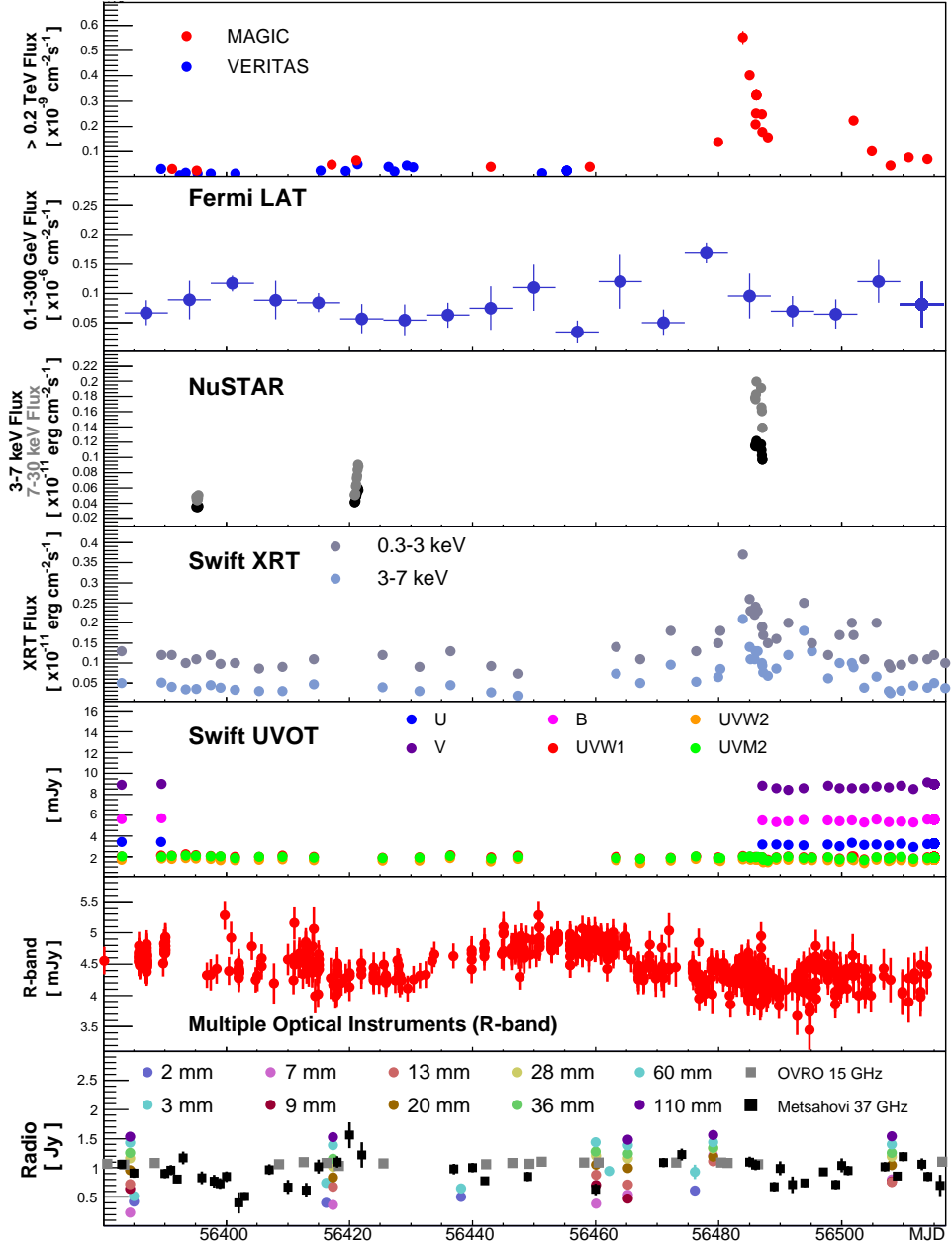}
\caption{The broadband light curves of Mrk\,501 from MJD 56380 to 56520.  The VHE data are shown with statistical error bars only. Optical data are corrected as described in Section 3.4. All radio light curve points for 2-110mm are provided by the F-Gamma consortium. }\label{fig1}
\end{figure*}

\begin{figure*}
\epsscale{0.9}
\center\includegraphics[width=5in]{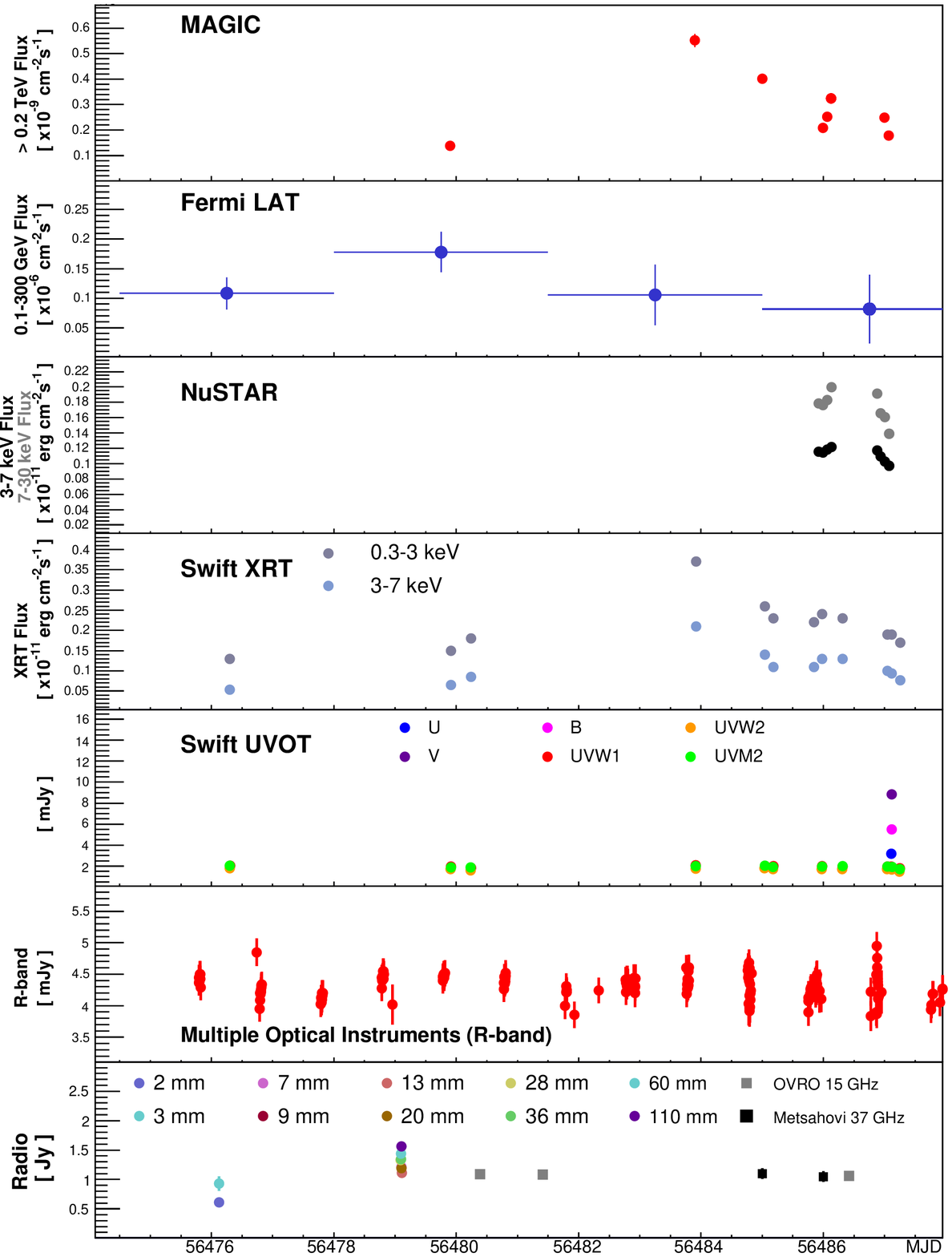}
\caption{The broadband light curve zoomed in to the period of the elevated X-ray and VHE gamma-ray state. }\label{fig1}
\end{figure*}

\subsubsection{VHE Results}

The full light curve of VHE observations from MAGIC and VERITAS is shown in Figure 5, with a zoom into the period of elevated flux in Figure 6.  The flux values are shown with statistical errors only.  The MAGIC and VERITAS observations of Mrk 501 in 2013 show the source in states which are consistent with the range of states observed in the past.  The observations of VERITAS, occurring primarily in the beginning of the campaign, detected the source in a 5-10\% Crab state, in agreement with the early MAGIC observations.  Later on in the campaign, MAGIC observed a flux elevated state of order $\sim2.5$ times the Crab flux.

\subsection{High-Energy Gamma Rays}
\textit{Fermi} LAT is a pair-conversion telescope sensitive to photons between 30 MeV and several hundred GeV \citep{atwood}.   Spectral analysis was completed for two periods contemporaneous with the \textit{NuSTAR} observations using the unbinned maximum-likelihood method implemented in the LAT \texttt{ScienceTools} software package version \texttt{v9r31p1}, which is available from the $Fermi$ Science Support Center.  The LAT data between MJD 56381 and MJD 56424 was used for comparison with the first two \textit{NuSTAR} exposures, while MJD 56471 to MJD 56499 was used for \textit{NuSTAR} exposures occurring during the elevated state.   

``Source" class events with energies above 100 MeV within a 12$^{\circ}$ radius of Mrk 501 with zenith angles $< 100^{\circ}$ and detected while the spacecraft was at a $<52^{\circ}$ rocking angle were used for this analysis.  All sources within the region of interest from the second \textit{Fermi} LAT catalog (2FGL, \citealt{2fgl}) are included in the model.  With indices held fixed, the normalizations of the components were allowed to vary freely during the spectral fitting, which was performed using the instrument response functions \texttt{P7REP\_SOURCE\_V15}.  The Galactic diffuse emission and an isotropic component, which is the sum of the extragalactic diffuse gamma-ray emission and the residual charged particle background, were modeled using the recommended files.\footnote{The files used were \texttt{gll\_iem\_v05\_rev1.fit} for the Galactic diffuse and \texttt{iso\_source\_v05.txt} for the isotropic diffuse component, both available at \tt http://fermi.gsfc.nasa.gov/ssc/data/access/lat/\\BackgroundModels.html}  The flux values were computed using an unbinned maximum likelihood analysis while fixing the spectral indices for the sources within the region of interest.  The systematic uncertainty of the LAT effective area is estimated as $10\%$ below 100\,MeV and decreasing linearly in Log(E) to 5\% between 316 MeV and 10 GeV.\footnote{\texttt{http://fermi.gsfc.nasa.gov/ssc/data/analysis/\\LAT\_caveats.html}}

The light curve for LAT observations of Mrk 501 was computed between MJD 56380 and 56520 in week-long bins (second panel from the top in Figure 5) and 3.5-day bins between MJD 56474 and 56488 (second panel from top of Figure 6).   Single day-binned light curve was also investigated, but no day within the time period provided a significant detection.  More specifically, no day provided a test statistic (TS; \citealt{mattox}) of greater than 9.  

During the first epoch (MJD 56381-56424), the spectral analysis of the LAT data shows the blazar had an integral flux of F$_{0.1-100 \rm{GeV}}$=(5.3$\pm$4.4)$\times10^{-8}$ph cm$^{-2}$s$^{-1}$, and an index of $\Gamma=2.0\pm0.3$.  Analysis of the second epoch (MJD 56471-56499) results in an integral flux of F$_{0.1-100 \rm{GeV}}$=(6.5$\pm$2.1)$\times10^{-8}$ph cm$^{-2}$s$^{-1}$ and index of  $\Gamma=1.7\pm0.1$.  These values are consistent with the average flux and index values calculated over the first 24 months of the science phase of the LAT mission and reported in the 2FGL catalog (F$_{0.1-100 \rm{GeV}}$=(4.8$\pm$1.9)$\times10^{-8}$ph cm$^{-2}$s$^{-1}$ and $\Gamma=1.74\pm0.03$; \citealt{2fgl}).

\subsection{\textit{Swift} X-Ray and UV Telescope Observations}
The XRT onboard \textit{Swift} \citep{gehrels} is a focusing X-ray telescope sensitive to photons with energies between 0.3 and 10 keV.  The \textit{Swift} satellite observed Mrk 501 59 times between 2013 January 1 and 2013 September 5 (MJD 56293 to 56540).  All XRT observations were carried out using the Windowed Timing (WT) readout mode.  The data set was first processed with the XRTDAS software package (v.2.9.0) developed at the ASI Science Data Center and distributed by HEASARC within the HEASoft package (v. 6.13). Event files were calibrated and cleaned with standard filtering criteria with the {\it xrtpipeline} task using the calibration files as available in the \textit{Swift} CALDB version 20140120.

The spectrum from each observation was extracted from the summed and cleaned event file.  Events for the spectral analysis were selected within a circle of 20 pixel ($\sim46 \arcsec$) radius, which encloses about 80\% of the \textit{Swift} XRT point spread function (PSF), centered on the source position. The background was extracted from a nearby circular region of 40 pixel radius.  The ancillary response files were generated with the {\it xrtmkarf} task, applying corrections for PSF losses and CCD defects using the cumulative exposure map. The latest response matrices (v.014) available in the \textit{Swift} CALDB were used. Before the spectral fitting, the 0.3-10 keV source energy spectra were binned to ensure a minimum of 20 counts per bin.  

The data were fit with an absorbed power-law model, with index $\Gamma$, as well as an absorbed log-parabolic model, where in both cases the neutral hydrogen column density was set at 1.55 $\times10^{20}$cm$^{-2}$, taken from \cite{kalberla}.  The summary of the XRT observations and spectral analysis results are provided in Table 4.   The light curve of the observations, including 0.3-3 keV and 3-7 keV integral flux bands, is shown in Figure 5, with a zoom into the period of elevated flux in Figure 6.  The 3-7 keV band is not traditionally quoted for \textit{Swift} XRT data, but is motivated by direct comparison to the 3-7 keV band computed for the \textit{NuSTAR} observations. 

Mrk 501 displays a relatively steady flux state until after MJD 56480, when the flux increases to (38.3$\pm$1.5) $\times10^{-11}$ ergs cm$^{-2}$s$^{-1}$  on MJD 56483 (corresponding to the day with the XRT count rate of 15 counts s$^{-1}$ which triggered MAGIC and \textit{NuSTAR} observations).  This high X-ray state was followed by a general drop in flux, continuing through the last XRT observation included in this work (2013 September 1; MJD 56540).

\input{SwiftXRT.tex}

The power-law fitted indices and 3-7 keV flux derived from the power-law fits are plotted in Figure 7 for all 59 observations.  The source clearly displays the harder-when-brighter trend found previously in other TeV blazars, such as Mrk 421 \citep{hardwhenbright}.  This behavior is similar to that displayed in the hard X-ray band 7-30 keV observed by \textit{NuSTAR} and shown in Figure 3.  Notably, the photon indices in the soft X-ray band are systematically harder than those observed by \textit{NuSTAR} in the 7-30 keV band.  The spectral index observed by \textit{Swift} XRT ($\Gamma$, determined at 1 keV) ranges between 1.4 and 2.2 (Figure 7) while the \textit{NuSTAR} index, determined at 10 keV, ranges from 2.1 to 2.4 (Figure 3).

\begin{figure}
\epsscale{0.9}
\includegraphics[width=3.4in]{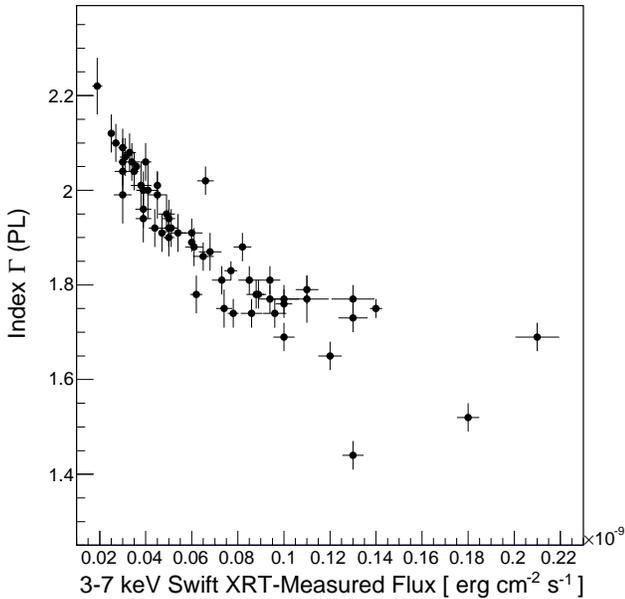}
\caption{The power-law index versus 3-7 keV flux values fit to the \textit{Swift} XRT observations of Mrk 501.}\label{fig1}
\end{figure}

Additionally, UV/optical observations were collected with the UVOT onboard \textit{Swift}.  These observations were carried out using the  ``filter of the day'', i.e. one of the six lenticular filters (V, B, U, UVW1, UVM2, and UVW2), unless otherwise specified in the ToO request, so images are not always available for all filters. There are 50 observations included in this Mrk 501 campaign, 18 of which included exposures in all filters while the remaining 32 observations contain UV imaging only. 

For each filter observation, we performed aperture photometry analysis using the standard UVOT software distributed within the HEAsoft 6.10.0 package and the calibration included in the latest release of CALDB.  Counts were extracted from apertures of 5\arcsec\, radius for all filters and converted to fluxes using the standard zero points from \cite{poole08}. The flux values were then de-reddened using the value of $E(B-V)$\,=\,0.017 \citep{Schafly2011} with $A_{\lambda}/E(B-V)$ ratios calculated for UVOT filters using the mean Galactic interstellar extinction curve from \cite{Fitzpatrick1999}.   No variability was detected to occur within single exposures in any filter.  The processing results were verified, checking for possible contamination from nearby objects falling within the background apertures.

\subsection{Optical}
Temporal coverage at optical frequencies was provided by various telescopes around the world, including the GASP-WEBT program
\citep[e.g.][]{Villata2008, Villata2009}. In particular, we report observations performed in the $R$-band from the following observatories: Crimean, Roque de los Muchachos (KVA), Lulin (SLT), Abastumani (70cm), Skinakas, Rozhen (60cm), Vidojevica (60cm), Perkins, Liverpool, St. Petersburg, West Mountain Observatory (WMO), the robotic telescope network AAVSOnet, the 60 cm and 1 m telescopes at the TUBITAK National Observatory (TUG T60 and TUG T100) and the Fred Lawrence Whipple Observatory (FLWO).    Host galaxy estimation for the $R$ filter is obtained from \cite{nilsson}, with apertures of 7.5\arcsec\, and 5\arcsec\,, used for the various instruments. Galactic extinction was accounted for according to the coefficients from \cite{Schafly2011}. The calibration stars reported in \citet{Villata1998} were used for calibration. 

Due to different filter spectral responses and analysis procedures of the various optical data sets (e.g. for signal and background extraction) in combination with the strong host galaxy contribution ($\sim$12 mJy for an aperture of 7.5\arcsec\, in the $R$-band), the reported fluxes required instrument-specific offsets of a few mJy. These offsets are introduced in order to align multi-instrumental light curves, and were determined using several of the GASP-WEBT instruments as reference, and scaling the other instruments using simultaneous observations. The required offsets for each instrument are as follows: Abastumani (70cm)=4.8 mJy; Skinakas=1.2 mJy; Rozhen (60cm)=-1.3 mJy; Vidojevica (60cm)=2.2 mJy; St.Petersburg=0.3 mJy; Perkins=0.6 mJy; Liverpool=0.6 mJy; AAVSOnet=-3.4 mJy; WMO= -0.7 mJy; TUG T60=0.5 mJy; TUG T100=-1.2 mJy. Additionally, a point-wise fluctuation of 0.2 mJy ($\sim$0.01mag) was added in quadrature to the statistical errors in order to account for potential differences of day-to-day observations within single instruments.  Within Figure 5, the $R$-band observations can be seen to remain fairly steady around 4.5 mJy.

\subsection{Radio}
\subsubsection{Mets\"ahovi}
The 14-m Mets\"ahovi Radio Observatory also participated in this multi-instrument campaign, as it has been doing since 2008. Mets\"ahovi observed Mrk\,501 every few days at 37 GHz.  Details of the observing strategy and data reduction can be found at \cite{terasranta}.  As can be seen in the bottom panel of Figure~5, there is evidence of a low level of variability at 37 GHz as observed by Mets\"ahovi.  This variability is quantified in terms of fractional variability (see Section 5.1).

\subsubsection{OVRO}
Regular 15\,GHz observations of Mrk\,501 were carried out using the OVRO 40-m telescope with a nominal bi-weekly cadence \citep{Richards2011}. The instrument consists of off-axis dual-beam optics and a cryogenic high electron mobility transistor low-noise amplifier with a 15\,GHz center frequency and 3\,GHz bandwidth. The two sky beams were Dicke-switched using the off-source beam as a reference, while the source was alternated between the two beams in an ON-ON mode to remove atmospheric and ground contamination. The total system noise temperature was about 52\,K. The typical noise level achieved in a 70-second observation was 3--4\,mJy. The flux density uncertainty includes an additional 2\% uncertainty mostly due to pointing errors, but does not include the systematic uncertainty in absolute calibration of about 5\%. Calibration was performed using a temperature-stable diode noise source to remove receiver gain drifts; the flux density scale is derived from observations of 3C\,286 assuming the \cite{Baars1977} value of 3.44\,Jy at 15\,GHz. Details of the reduction and calibration procedure can be found in \citet{Richards2011}.

\subsubsection{F-Gamma}
The cm/mm radio light curves of Mrk\,501 were obtained within the framework of a {\sl Fermi}-related monitoring program of gamma-ray blazars (F-Gamma program; \cite{fuhrmann,angelakis}). The millimeter observations were closely  coordinated with the more general flux monitoring conducted by IRAM, and data from both programs are included here. The overall frequency range spans from 2.64\,GHz  to 142\,GHz using the Effelsberg 100-m and IRAM 30-m telescopes.

The Effelsberg measurements were conducted with the secondary focus heterodyne receivers at 2.64, 4.85, 8.35, 10.45, 14.60, 23.05, 32.00 and 43.00\,GHz. The observations were performed quasi-simultaneously with cross-scans; that is, slewing over the source position, in azimuth and elevation direction with an adaptive number of sub-scans for reaching the desired sensitivity (for details, see Fuhrmann et al. 2008; Angelakis et al. 2008). Subsequently, pointing offset correction, gain correction, atmospheric opacity correction and sensitivity correction were applied to the data.

The IRAM 30-m observations were carried out with calibrated cross-scans using the Eight MIxer Receiver (EMIR) horizontal and vertical polarization receivers operating at 86.2 and 142.3\,GHz.  The opacity-corrected intensities were converted to the standard temperature scale and finally corrected for small remaining pointing offsets and systematic gain-elevation effects. The conversion to the standard flux density scale was done using the instantaneous  conversion factors derived from frequently observed primary (Mars, Uranus) and secondary (W3(OH), K3-50A, NGC\,7027) calibrators.

\section{Simultaneous \textit{NuSTAR} and \textit{Swift} Exposures}
Since Mrk 501 is highly variable,  detailed inferences regarding the broadband SED and its temporal evolution require simultaneous observations of multiple bands.  In particular, for the determination of the low-energy peak $E_{\rm{syn}}$, and the flux at $E_{\rm{syn}}$, $F(E_{\rm{syn}})$, \textit{Swift} XRT and \textit{NuSTAR} observations must be simultaneous.  There are five periods within the campaign for Mrk 501 where the observations by \textit{NuSTAR} and \textit{Swift} occurred within one hour of each other.  The \textit{Swift} exposure IDs for these quasi-simultaneous periods are summarized in Table 4.  For Mrk 501, $E_{\rm{syn}}$ is located in the X-ray band and can be determined reliably (except for the first \textit{NuSTAR} observation where $E_{\rm{syn}}$ is $\le0.85$ keV) since there is no evidence of X-ray variability of Mrk 501 on a time scale shorter than a \textit{NuSTAR} orbit ($\sim90$ minutes).

As a precursor to the joint fitting of XRT and \textit{NuSTAR} data, we confirm agreement between the 3-7 keV flux values derived from the \textit{Swift} XRT and \textit{NuSTAR} fitted models.  There is a residual discrepancy (not a uniform offset) at the level of $< 10\%$.  Using XSPEC, we performed simultaneous fitting to the datasets using the absorbed log-parabolic model as done in Section 2 for the \textit{NuSTAR} data alone.  During the fitting process, we allowed the normalizations of the data sets to vary, but required the same spectral shape parameters.  A representative plot of the simultaneous fit for XRT and \textit{NuSTAR} data collected on MJD 56485 is provided in Figure 8.  The model spectrum is shown as a solid line in Figure 8.  The agreement between XRT and \textit{NuSTAR} was studied and found to be within the calibration uncertainties\footnote{\tt{http://heasarc.gsfc.nasa.gov/docs/heasarc/caldb/swift\\/docs/xrt/SWIFT-XRT-CALDB-09\_v18.pdf}}.

\begin{figure*}
\epsscale{0.9}
\includegraphics[width=4.8in,angle=-90]{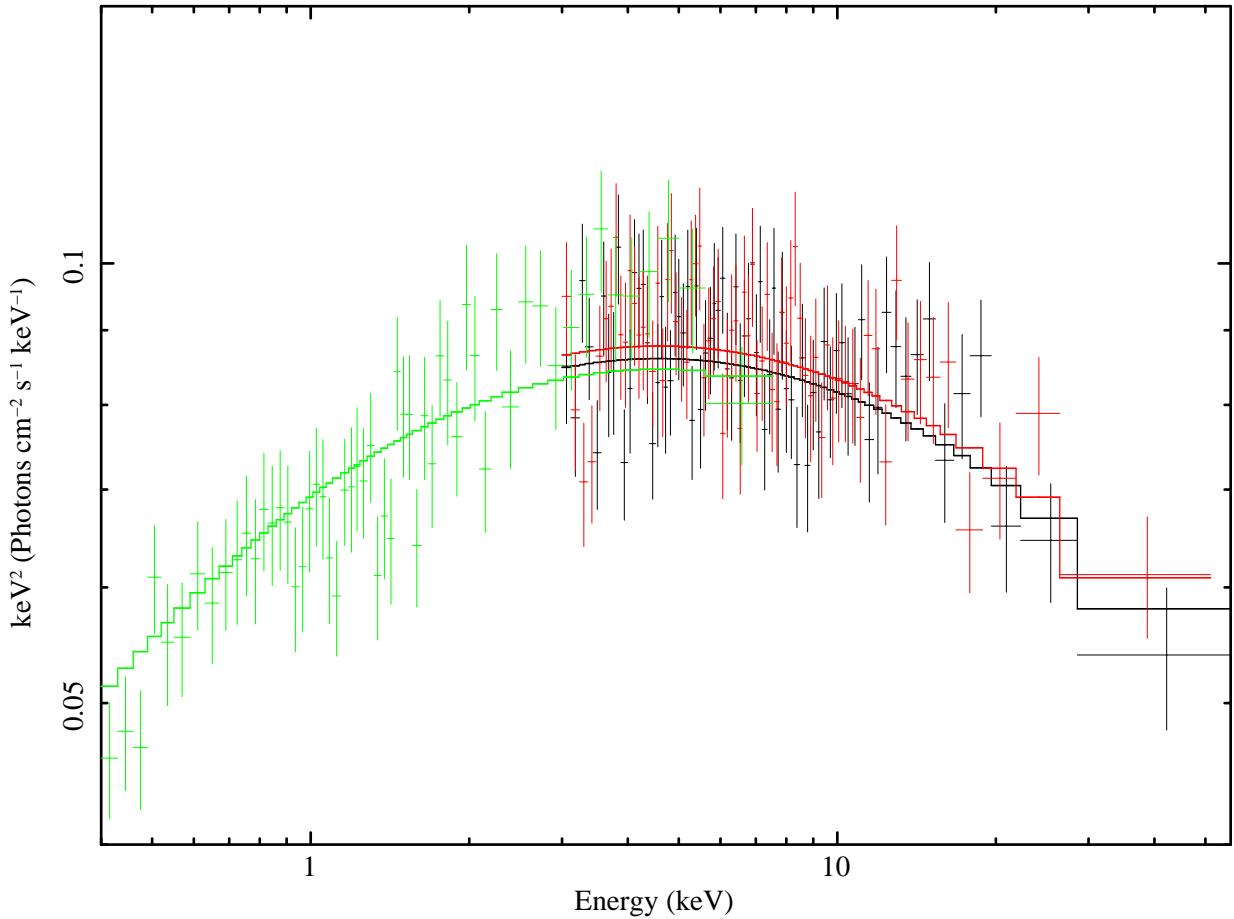}
\caption{Example of a broadband X-ray spectrum of Mrk 501 in the
crucial region where the synchrotron peak (in the $E \times F(E)$
representation) is located.  The spectra result from a simultaneous observation
with \textit{Swift} (green) and \textit{NuSTAR} (FPMA: red, FPMB: black) on 2013 July 12-13.  The spectral fit used a log-parabolic model
(see the text) with Galactic column density of $1.55 \times 10^{20}$ cm$^{-2}$.  For the purpose of illustrating the intrinsic spectrum of the source, the solid lines $\-$
which represent the fit to the \textit{Swift} and \textit{NuSTAR} data $\-$ show the spectrum {\sl before} the Galactic absorption.
The normalizations of the \textit{Swift} and \textit{NuSTAR} data were allowed to be free, and
the offset between them was less than 10\%, thus illustrating generally good cross-calibration of the two instruments.
}
\end{figure*}

For the determination of the spectral parameters characterizing the synchrotron peak (namely the energy $E_{\rm syn}$ and $F(E_{\rm syn}$)) with the simultaneous \textit{NuSTAR} and \textit{Swift} XRT observations, we apply the log-parabolic model modified by the photoelectric absorption due to our Galaxy, with a (fixed) neutral hydrogen column density of $1.55 \times 10^{20}$ cm$^{-2}$, taken from \cite{kalberla}. The procedure to search for $E_{\rm syn}$ involves the variation of the ``normalization energy" parameter (in the $\tt{logpar}$ model in XSPEC) until the local index $\Gamma$ returns a value of 2 --- then $E_{\rm syn}$ corresponds to the peak in the $E \times F(E)$ representation. This procedure correctly accounts for the effect of the soft X-ray absorption by Galactic column density as the absorption is included in the model fitted to the data.  For the determination of the error on $E_{\rm syn}$, we freeze the ``local index" --- defined at energy $E_{\rm syn}$ --- to a value of 2, and then step the value of $E_{\rm syn}$ keeping all other parameters free.  We then search for the value of the $E'_{\rm syn}$ which corresponds to the departure of $\chi^{2}$   from the minimum by $\Delta\chi^{2} = 2.7$.  The error quoted is the difference between $E_{\rm syn}$  and $E'_{\rm syn}$.  The $E_{\rm syn}$ and curvature parameters ($\beta$) for each of the simultaneous data sets are summarized in Table 5.  We quote the value of $F(E_{\rm syn})$ inferred from the \textit{NuSTAR} module FPMA (Focal Plane Module A).

The combination of \textit{Swift} XRT and \textit{NuSTAR} observations provides an unprecedented view of the synchrotron peak variability.  From Table 5, it is evident that the synchrotron peak moves by a factor of about ten during this campaign, with the highest synchrotron peak occurring during the elevated X-ray and gamma-ray state.

\begin{deluxetable*}{ccccccc}
\tablecolumns{7}
\tablewidth{0pc}
\scriptsize
\tablecaption{Fitting results for \textit{Swift} XRT and \textit{NuSTAR} simultaneous observations.  The data were simultaneously fit with a log-parabolic function.  }
\scriptsize
\tablehead{
\colhead{Observation } & \colhead{Date} & \colhead{Orbit} & \colhead{$E_{\rm syn}$}       & \colhead{$F(E_{\rm syn})$    } & \colhead{Curvature}  &\colhead{$\chi^2$/DOF}\\
\colhead{ID}& \colhead{[MJD]} & \colhead{Number} & \colhead{ [keV]  }       & \colhead{[$\times 10^{-11}$ ergs cm$^{-2}$s$^{-1}$]      } & \colhead{$\beta$}  &\colhead{}
 }
\startdata
60002024002 & 56395.1&  1&$<$0.85 &4.1&0.061&669/673\\
60002024006 & 56485.9 & 1&4.9$\pm$0.7&13.8&0.21&596/577\\
60002024006 & 56486.0 & 2&5.1$\pm$0.9&13.7&0.22&697/715\\
60002024006 & 56486.2 & 4&7.0$\pm$0.8&14.6&0.2&877/848\\
60002024008 & 56487.1& 4&3.3$\pm$0.9&11.2&0.17&832/851\\
\enddata
\end{deluxetable*}

\section{Variability}
\subsection{Fractional Variability}
In order to quantify the broadband variations we utilize the fractional variability, $F_{\rm var}$.  We follow the description given in \cite{vaughan}, where 
$F_{\rm var}$ is calculated as:

\begin{equation}
F_{\rm var} = \sqrt{\frac{S^2 -  \langle \sigma^2 \rangle}{\langle F_{\gamma}\rangle^2}} 
\end{equation}

\noindent where $  \langle F_{\gamma}\rangle $ is the average photon flux, $S$ is the standard deviation of the flux measurements, 
and $\langle\sigma^2\rangle$ is the mean squared error of the measurement. 

\begin{figure}
\epsscale{0.9}
\includegraphics[width=3.5in]{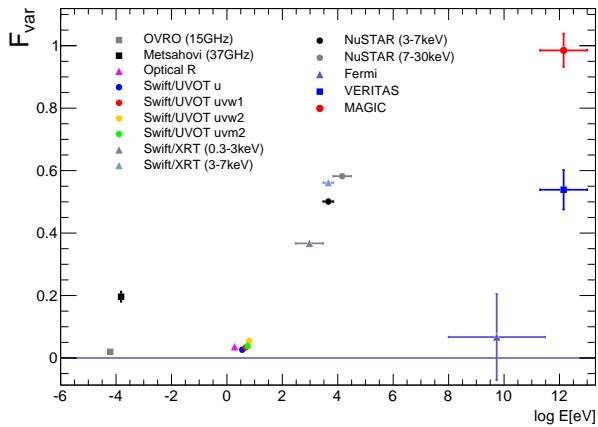}
\caption{The fractional variability ($F_{\rm var}$) calculated for each instrument separately.  }\label{Fvar}
\end{figure}

$F_{\rm var}$ was determined for the temporal binning and sampling presented in Figure 5 and Table 3 (for MJD 56485 and 56486, the bold lines in Table~3 are used).  The value of $F_{\rm var}$ is known to be dependent on sampling and should be interpreted with caution.  For example, a well sampled light curve with small temporal bins will allow us to probe the variability on small timescales (e.g. \textit{NuSTAR}), which could be hidden if the variability is computed with fluxes obtained with relatively coarse temporal bins (e.g. $Fermi$ LAT). 

The fractional variability for each band (from 15 GHz radio through VHE) is shown in Figure \ref{Fvar}.  For the period of observations covered in this work, the fractional variability shows a double-peaked shape with the highest variability in the X-ray and VHE bands. A similar broadband variability pattern has recently been reported for Mrk\,501 \citep{doert, mrk5012015}, for Mrk\, 421 \citep{magic421march,balokovic} and for other high-synchrotron-peaked blazars in, for example, \cite{2001}.   This double-peaked shape of $F_{\rm var}$ from radio through VHE can be interpreted as resulting from a correlation between the synchrotron and inverse-Compton peaks. 

$F_{\rm var}$ is below $\sim$5\% at 15 GHz and optical/UV frequencies, while at 37 GHz the fractional variability is $\sim$20\%. The relatively high fractional variability at 37 GHz is not produced by any single flaring event, but rather by a consistent flickering in the radio flux. Such flickering is not typically observed in blazars, but has been reported for Mrk\,501 in \cite{mrk5012015}.  At X-ray frequencies, $F_{\rm var}$ gradually increases with energy, reaching the largest value ($\sim$0.6) in the 7-30 keV band measured by \textit{NuSTAR}.  The $F_{\rm var}$ computed for the \textit{Swift} XRT 3-7 keV observations is higher than for the \textit{NuSTAR} 3-7 keV fluxes due to the larger temporal coverage of the \textit{Swift} observations, allowing for observation of Mrk\,501 during high activity levels that were not observed with \textit{NuSTAR}. 

The \textit{Swift} XRT $F_{\rm var}$ for Mrk\,501 published in \cite{stroh} was 0.15 or 0.18, depending on the timescale used for calculation, illustrating that the value of $F_{\rm var}$ is dependent on sampling.  In \cite{abdoMrk501}, RXTE-ASM (2- 10 keV) and Swift BAT (15-50 keV) show $F_{\rm var}$ values between 0.2 and 0.3, although it should be noted that due to the limited sensitivity of RXTE-ASM and \textit{Swift} BAT (in comparison with \textit{Swift} XRT and \textit{NuSTAR}), the variability was studied on timescales larger than 30 days.

\subsection{Cross Correlations}
Cross-correlations between the different energy bands were studied with the Discrete Correlation Function (DCF) described in \cite{edelson}. 
The DCF method can be applied to unevenly sampled data, and no interpolation of the data points is necessary. 
Also, the errors in the individual flux measurements are naturally taken into account when calculating the DCF. 
One important caveat, however, is that the resulting DCF versus time lag relation is not continuous, and hence the results should only be interpreted with a reasonable balance between the time resolution and the accuracy of the DCF values. 
It is also important to only consider instruments with similar time coverage. 
In this study, we considered all the energy bands with a non-zero fractional variability.
Among the \textit{Swift} UVOT data, only the UVW2 filter was checked, as it is the filter which has the best time coverage across the \textit{Swift} UVOT observations and also is least contaminated by the host galaxy light. 
For a better time coverage, MAGIC and VERITAS data points are combined to make a single data set as the VHE band. 

A significant correlation in the DCF was seen only between the VHE data and the 0.3-3 keV and the 3-7 keV \textit{Swift} XRT bands. 
For both of the combinations, the largest correlation is seen with a time lag of $0\pm 1.5$ days.  This result does not change if the binning of 3 days is altered. 
Note that the \textit{NuSTAR} observations covered a relatively short period with a dense sampling, thus we did not see any significant correlation between \textit{NuSTAR} and any other band. 
Since the observations of \textit{Swift} XRT and \textit{NuSTAR} were made simultaneously (within a few hours) with the VHE observations, correlations between the X-ray and the VHE observations were investigated in more detail (see Section 5.3). 
R

\subsection{X-ray/VHE Correlation}
The light curve of the broadband observations is shown in Figure 5, with a zoom of the period showing an elevated  X-ray and VHE state in Figure 6.  The VERITAS and MAGIC flux points within the light curve are shown with statistical errors only.  Correlation studies using the VHE flux values are completed with statistical and systematic errors included, as described below.    The radio, optical and UV observations show relatively steady flux over the campaign period, while the largest amplitude of variability can be seen in the X-ray and VHE gamma-ray bands.  An elevated state in both the X-ray and VHE bands can be seen to occur on MJD 56483 (\textit{Swift} Observation ID 00030793232 in Table 4).  Zooming in on this epoch (Figure 6), shows that the \textit{NuSTAR} observations occurring on MJD 56485 and 56486 occurred after the highest state observed by MAGIC and \textit{Swift}.  The XRT observations show an elevated X-ray flux in both the 0.3-3 and 3-7 keV bands on MJD 56483.

\begin{figure*}
\epsscale{0.9}
\center\includegraphics[width=6in]{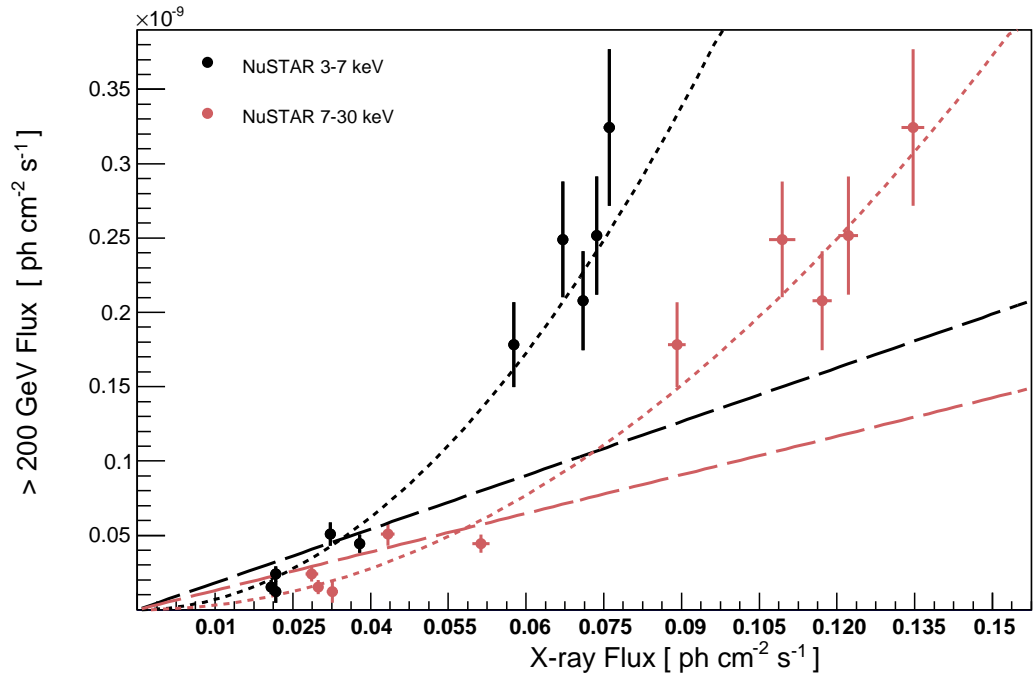}
\caption{\textit{NuSTAR} X-ray photon flux versus simultaneous $>200$ GeV flux from MAGIC and VERITAS.  The dotted lines show quadratic fits to the data, while the dashed lines show linear fits to the 3-7 keV and 7-30 keV bands. }\label{fig1}
\end{figure*}

A comparison between the \textit{NuSTAR}-observed X-ray photon flux values (derived from XSPEC) in the 3-7 and 7-30 keV bands and the epochs of simultaneous VHE observations is shown in Figure 10.   During this campaign, 10 observations occurred within one hour between either \textit{NuSTAR} and MAGIC (7 observations) or \textit{NuSTAR} and VERITAS (3 observations).  The simultaneous X-ray and VHE data, where the VHE data include both statistical and systematic errors, were fit with both a linear and a quadratic function.   

Within the one-zone SSC emission paradigm, there is a physical motivation for a quadratic relationship between the X-ray and VHE flux values \citep{marscher}.  More specifically, the inverse-Compton flux depends not only on the density of photons, but also on the density of the electron population producing those photons. If, however, the particle population is energetic enough for the inverse-Compton scattering to occur in the Klein-Nishina regime, the relationship between the X-ray and VHE fluxes can be complex and will depend in detail on the energy bands considered, the particle energy loss mechanisms and the magnetic field evolution.  In particular, \cite{katarzynski} suggest that a roughly linear relationship may arise during the declining part of a flare when the emitting region expands adiabatically, leading to a decrease of both the particle number density and the magnetic field strength.



A quadratic relationship provides a better fit than the linear fit for the 3-7 keV flux values measured simultaneously by \textit{NuSTAR}, with $\chi^2$ of 11.4 and 87.3, respectively, for 9 DOF.  The 3-7 keV flux and the $> 200$ GeV flux are highly correlated, with a Pearson correlation coefficient ($r$) of 0.974.  Similarly, for the 7-30 keV band, the quadratic relation fits the data better than the linear relation, with $\chi^2$ of 17.5 and 79.1, respectively, for 9 DOF.  The $r$-value for the 7-30 keV flux and the $>200$ GeV flux is 0.979.


\begin{figure*}
\epsscale{0.9}
\center\includegraphics[width=6in]{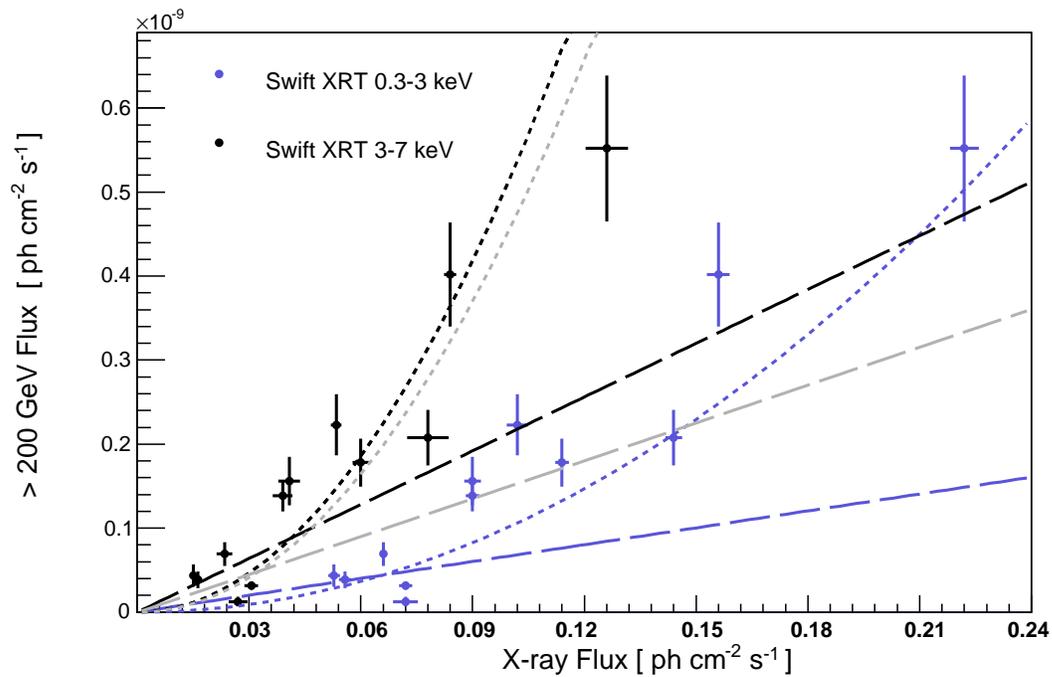}
\caption{\textit{Swift} 0.3-3 and 3-7 keV X-ray photon flux values versus simultaneously measured $>200$ GeV flux from MAGIC and VERITAS.  The black and blue dotted lines show quadratic fits to the 3-7 keV and 0.3-3 keV data, respectively, while the black and blue dashed lines show linear fits to the 0.3-3 keV and 3-7 keV bands, respectively.  For completeness, we also compare the linear and quadratic fits of the simultaneous 3-7 keV \textit{NuSTAR} and $>200$ GeV flux from MAGIC and VERITAS summarized in Figure 10 (light grey dashed and dotted line). }\label{fig1}
\end{figure*}

A comparison between the \textit{Swift}-observed X-ray photon flux values (derived from XSPEC) in the 0.3-3 and 3-7 keV bands and the epochs of simultaneous VHE observations is shown in Figure 11.  These data are not simultaneous with the \textit{NuSTAR} observations shown in Figure 10 and therefore the results cannot be directly compared.  During this campaign, 12 absolutely simultaneous observations occurred between \textit{Swift} and MAGIC (10) and \textit{Swift} and VERITAS (2), shown in Figure 11.  
Similarly as done for the \textit{NuSTAR} bands, the simultaneous \textit{Swift} X-ray and VHE data were fit with both a linear and a quadratic function with an offset fixed to zero.  For the 0.3-3 keV flux values measured simultaneously by \textit{Swift}, a quadratic relationship provides a better fit than the linear fit, with $\chi^2$ of 81.8 and 162.0, respectively, for 11 DOF.  The 0.3-3 keV flux and the $> 200$ GeV flux are highly correlated, with a Pearson correlation coefficient ($r$) of 0.958.  For the 3-7 keV band, the quadratic function fits the data better than the linear function, with $\chi^2$ of 58.0 and 114.0, respectively, for 11 DOF.  The $r$-value for the 3-7 keV flux as measured with \textit{Swift} and the $>200$ GeV flux is 0.954.   


\section{Modeling the Broadband Spectral Energy Distribution}
Previous MWL campaigns on Mrk 501 have been sufficiently characterized with a one-zone SSC model \citep{mrk501past,abdoMrk501}, although there are a few notable instances where a one-zone SSC model was found not to be appropriate for the broadband emission \citep{pian, kataoka}.  In this study we decided to use the simplest approach, which is provided by a leptonic model with a single emitting region.  The broadband spectral data were modeled with an equilibrium version of the single-zone SSC model from \cite{bottcherchiang} and \cite{bottcher2013}. This model has been used to describe the broadband emission from various other VHE-detected blazars \citep[e.g.][]{wcom,3C66A,RXJ0648,B21215}.   

Within this equilibrium model, the emission originates from a spherical region of relativistic leptons with radius $R$.  This emission region moves down the jet with a Lorentz factor $\Gamma$.  We set the Doppler factor $\delta$ to 15 for all model representations.   Notably, it has been shown that when using least-squares fitting of emission models to broadband data of Mrk\,501, the Doppler factor can vary widely from state to state \citep{mankuzhiyil}.  We do not complete least-squares fitting in this work and instead choose to fix the Doppler factor to 15 for the representation of all states, limiting the number of free parameters of the SSC model.  The Doppler factor of 15 is similar to the Doppler factor used in previous studies of Mrk 501 \citep{mrk501past, abdoMrk501,mankuzhiyil}.  In order to reduce the number of free parameters, the jet axis is aligned toward the line of sight with the critical angle $\theta=3.8^{\circ}$.  At the critical angle, the jet Lorentz factor is equal to the Doppler factor ($\Gamma=\delta$). 

Within this emission model, relativistic leptons are injected into this emission region continuously according to a power-law distribution $Q=Q_{0}\gamma^{-q}$ between $\gamma_{\rm min}$ and $\gamma_{\rm max}$. The injected population of particles is allowed to cool.  The simulation accounts for synchrotron emission due to a tangled magnetic field $B_0$, Compton up-scattering of synchrotron photons, $\gamma\,\gamma$ absorption and the corresponding pair production rates (via the general solution in \citealt{bottcherschlickeiser}).  The cooling of the injected electrons is dominated by radiative losses, which are balanced by injection and particle escape from the system.  This particle escape is characterized with an escape efficiency factor $\eta$=100, where $t_{\rm esc}=\eta$R/c.  The use of $\eta$=100 is motivated by success in representing SEDs of TeV blazars in previous studies using the same model (e.g. \citealt{B21215}).    The electron cooling rates and photon emissivity and opacity are calculated using similar routines of the code for jet radiation transfer described in \cite{bottchermauseschlickeiser}.  Together, the particle injection, cooling and escape mechanisms lead to an equilibrium particle population.

A key result of the equilibrium that occurs between continual particle injection, particle escape and radiative cooling is a break in the electron distribution $\gamma_b$ (referred to as $\gamma_c$ within \citealt{bottcher2013}), where $t_{esc}=t_{cool}(\gamma_b)$.  As described in Equations (1) and (2) of \cite{bottcher2013}, if $\gamma_b$ is smaller than $\gamma_{min}$, the system will be in a fast cooling regime.   If $\gamma_b$ is greater than $\gamma_{min}$, the system will be in a slow cooling regime.   Within the fast cooling regime, the equilibrium particle distribution is a broken power law, with an index of 2 for particles with Lorentz factors less than $\gamma_{min}$, and an index of ($q$+1) for Lorentz factors above $\gamma_{min}$.  In the slow cooling regime, the resulting broken power law of the equilibrium particle distribution is equal to the injected spectrum ($q$) for particles with Lorentz factor below $\gamma_b$, and $(q+1)$ above $\gamma_b$.  
It is known that a hard injected electron spectrum would lead to a small amount of pile-up, followed by a smooth cut-off toward the high-energy end of the distribution (for details, see, e. g.  \citealt{kardeshev} and \citealt{stawarz}).  More specifically, the equilibrium electron spectrum slightly deviates from the ($q+1$) approximation at the high-energy end ($\gamma \sim \gamma_{\rm max}$) due to pile-up effects that increase as the injected spectrum becomes harder (i.e. $q<1.5$).    Notably, although scattering in the KN regime is appropriately accounted for within the SSC model, neither the pile-up at the highest energy nor the energy loss (Compton cooling) of the electrons participating in scattering within the KN regime is accounted for within the model.  The two aforementioned effects, however, are expected to result in a negligible deviation of the equilibrium electron spectrum from the approximated index of $q+1$.

$L_e$ is the kinetic power in the relativistic electrons and $L_B$ is the power in the Poynting flux carried by the magnetic field of the equilibrium particle distribution.  The $L_e$ and $L_B$ parameters allow the calculation of the equipartition parameter $L_B/L_e$.  
 A state with an equipartition near unity minimizes the total (magnetic field +
    particle) energy requirement to produce a given synchrotron
    flux. Therefore, from an energetics point of view a situation
    near equipartition is usually favored. If the jet is powered 
    by a Blandford-Znajek type mechanism, it is expected to be
    initially Poynting-flux dominated, and this luminosity is
    then (through an unknown mechanism, possibly magnetic 
    reconnection) converted partially into particle energy.
    This conversion is expected to stop at an approximately
    equipartition situation as an equilibrium is reached between the conversion of magnetic energy to particle energy, and vice versa (via turbulent
charged-particle motion generating small-scale, turbulent magnetic fields).  For examples of blazar modeling based on equipartition, see \cite{matteo1,matteo2}.   Alternatively, a sub-equipartition
    magnetic field may be expected in an MHD-driven, initially
    particle-dominated jet, where magnetic fields could be self-generated
    (amplified) by, e.g., shocks. The sub-equipartition magnetic fields
    that are often found in blazar SED modeling might therefore
    favor this latter scenario.  
    Sub-equipartition states are a common result in the application of single-zone SSC emission scenarios to VHE blazars, as in \cite{aliu2012a, aliu2012b,RXJ0648,wcom,acciari2009b,acciari2009c,acciari2008b} and \cite{abdo2011c}.   

\begin{figure*}
\epsscale{0.9}
\center\includegraphics[width=6.3in]{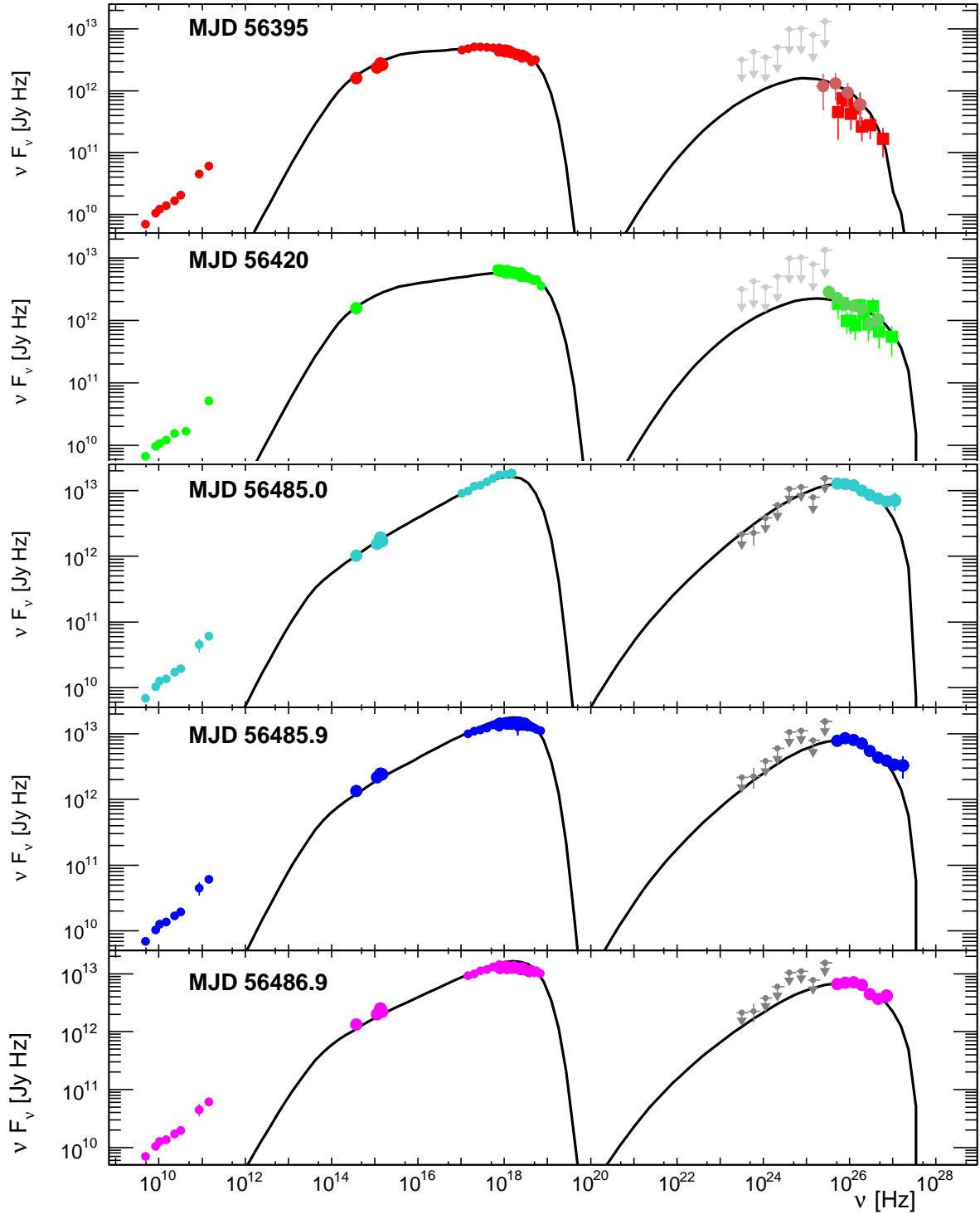}
\caption{Observed broadband SEDs of Mrk\,501 on each of the days where \textit{NuSTAR} observations occurred (red, green, blue and pink data). Additionally we include observations from MJD 56485.0 (turquoise, center panel), which show the SED one day after the most elevated flux state observed during this campaign.  The broadband data are represented with a single-zone SSC model (solid line), with the model parameters summarized in Table 6.  The $Fermi$ LAT limits shown in the top two panels are taken from analysis of data between MJD 56381 and 56424, while the bottom three panels show \textit{Fermi} results produced from analysis of data between MJD 56471 and 56499. }\label{fig1}
\end{figure*}

The broadband data and model representations for five days from the MWL observation campaign are shown in Figure 12.  The flux resulting from the model simulation (solid line) is corrected for absorption by interaction with the extragalactic background light (EBL) for the redshift of $z=0.03$, assuming the EBL model outlined in \cite{dominguez}.  The model thus represents the observed VHE emission as opposed to the intrinsic VHE emission.  When applying the model to the data, the radio flux is likely to include a significant portion of extended radio emission and is therefore taken as an upper limit, as done in \cite{abdoMrk501}.

\begin{deluxetable*}{ccccccc}
\tablecolumns{7}
\tablewidth{0pc}
\scriptsize
\tablecaption{Single-zone SSC model parameter values (see Section 6 for overview of model and parameters).  Model representations are shown along with the broadband data in Figure 12.}
\scriptsize
\tablehead{
\colhead{Parameter} & \colhead{MJD 56395    }       & \colhead{MJD 56420      } & \colhead{MJD 56485.0}& \colhead{MJD 56485.9} &\colhead{MJD 56486.9}  }
\startdata
$\gamma_{min}$ [$\times10^4$]&2.0&2.1&2.0&2.0&2.0\\
$\gamma_{max}$ [$\times10^6$] &1.0&1.4&1.4&1.7&1.4\\
$\gamma_{break}$ [$\times10^3$]&4.1&4.6&2.8&3.3&3.4\\
$q$&1.9&1.8&1.3&1.3&1.3\\
$\eta$&100&100&100&100&100\\
$\delta$ &15 &15 & 15 & 15 & 15\\
B$_0$ [G]&0.06&0.05&0.03&0.03&0.03\\
$\Gamma$&15&15&15&15&15\\
R [$\times10^{15}$ cm ]&7.0&7.0&5.0&7.0&7.0\\
$\theta$ [degrees]&3.8&3.8&3.8&3.8&3.8\\
$t_{var}$ [hr]&4.3&4.3&3.1&4.3&4.3\\
$L_e$ [$\times10^{42}$ erg s$^{-1}$]&9&12&36&28&26\\
$\epsilon$=$L_B/L_e$&1.8$\times10^{-2}$&6.1$\times10^{-2}$&5.3$\times10^{-4}$&1.3$\times10^{-3}$&1.4$\times10^{-3}$\\
\enddata
\end{deluxetable*}

The parameters used to represent the data with the equilibrium model are summarized in Table 6. The data in this work are represented with the emission model within the fast cooling regime, where the emitting equilibrium particle population follows $n(e)\propto\gamma^{-2}$ for $\gamma_b<\gamma<\gamma_{min}$ and $n(e)\propto\gamma^{-(q+1)}$ for $\gamma_{min}<\gamma<\gamma_{max}$.  A particle population with an injection index of $q=1.8-1.9$ provides a reasonable representation of the synchrotron emission on MJD 56395 (red; top panel) and 56420 (green; second panel from top).  There are no \textit{Swift} data for observations on MJD 56420.  Each of these epochs (MJD 56395 and 56420) can be sufficiently described with similar parameters, although the SED on MJD 56420 requires a slightly more energetic electron population and lower magnetic field to account for the marginally elevated X-ray and VHE emission as compared to what is observed on MJD 56395.

Although the highest VHE gamma-ray ($\ge200$ GeV) flux during this campaign was observed by MAGIC on MJD 56484, a reliable spectrum from that MAGIC observation could not be reconstructed due to the presence of Calima and the lack of LIDAR data to correct for it.  \textit{Swift} XRT also recorded the highest X-ray flux in its observation on the same day.  On the other hand, there are sufficient broadband data to model the SED on MJD 56485.0 (turquoise; middle panel Figure 12), which is less than one day later than the MAGIC and \textit{Swift} observation of the highest fluxes occurred.

The light curve in Figure 5 shows that Mrk 501 displayed relatively steady emission in each band between MJD 56420 and the elevated state observed by \textit{Swift} and MAGIC on MJD 56484.  In moving from the relatively quiescent SED on MJD 56420 to the elevated state observed on MJD 56485, a hardening of the injection spectrum is required ($q$=1.3) to match the X-ray spectrum observed by \textit{Swift} XRT.  With the injection index responsible for the hardness of the synchrotron emission at X-ray energies, the frequency at which the synchrotron emission peaks, is related to the spectrum of the injected particle population, and the magnetic field ($B_0$).  When moving from the state on MJD 56420 to 56485.0, the strength of the magnetic field decreases, moving the peak of the synchrotron emission to lower energies. The decrease of the synchrotron flux resulting from a lower magnetic field is counteracted with an increase of particle luminosity $L_e$.   Finally, to match the relative magnitudes of the synchrotron and inverse-Compton peak fluxes, the electron and photon density of the emission region was increased with a decrease of the emission region size.  The decrease of the emission region size to 5.0 $\times10^{15}$ cm on MJD 56485.0 provides a higher inverse-Compton flux while maintaining the synchrotron flux. 

Following \cite{blumenthal}, the regime at which the up-scattering is occurring can be estimated (in the observer frame) according to $4\,h\nu_{\rm syn\,pk}\gamma/\delta m_e\,c^2$, where $\gamma$ represents the energy of the electrons up-scattering $\nu_{\rm syn\,pk}$ photons.  If this quantity is less than 1, the inverse-Compton emission is occurring within the Thomson regime, while if it is greater than 1, the emission is occurring in the KN limit.  With $\nu_{\rm syn\,pk}$ at approximately 5 keV, $4\,h\nu_{\rm syn\,pk}\gamma_{\rm min}/\delta m_e\,c^2\sim25$, indicating that, according to the model applied within this work, the inverse-Compton scattering of the photons near the synchrotron peak is far into the KN regime.  We note that this is not necessarily in conflict with the quadratic relationship between the simultaneous X-ray and VHE flux measurements, but it implies a reasonably steady value of magnetic field which is supported by our SSC models;  see Table 6.  For a more extensive discussion, see \cite{katarzynski}.


The SEDs on the days MJD 56485.9 and 56486.9 are similar to MJD 56485.0.  All model representations explored here result in emission scenarios which are heavily matter dominated (far below equipartition), where the majority of the energy is distributed within the particle population instead of in the magnetic field.    Notably, even a single-zone SSC model is difficult to constrain, and the solutions presented here are not applied with the intent of constraining parameter space, but instead to just show that a reasonable representation of the data is possible.  There are additional models (e.g. multi-zone or hadronic models) which might alternatively be used to describe the broadband emission from Mrk 501 during these epochs (e.g. \citealt{tavecchio, magic421march}).  However, these models have twice as many free parameters as single-zone leptonic models and, in this particular case, there are not strong constraints from MWL flux evolution correlations that point to the necessity of such models.

\section{Discussion and Conclusions}
The inclusion of the hard X-ray telescope \textit{NuSTAR} in this observational campaign has provided unprecedented insight into the temporal evolution of the 3-30 keV X-rays emitted by Mrk\,501.  Before this campaign, Mrk 501 had not been observed to display hard X-ray variability on timescales of $\sim$7-hours.  The fractional variability of Mrk\,501 observed during this campaign was highly significant for the \textit{NuSTAR} 7-30 keV band ($F_{\rm var}$=0.6).

Investigation of the DCF allows insight into possible leads or lags between the low (0.3-3 keV) and high (3-7 keV) X-ray and VHE emission. The variability between these two bands shows evidence for a zero day lag.  Correlation between the X-ray and VHE bands is further supported by the correlated variability inferred from the Pearson coefficients of 0.958 and 0.954 for simultaneous observations (occurring within one hour), respectively.   Correlation is also found between the \textit{NuSTAR} X-ray flux values and the simultaneous $>$ 200 GeV flux values (with observations occurring within one hour), with Pearson coefficients of 0.974 and 0.979 for the 3-7 keV and 7-30 keV bands, respectively.  

Correlation of variability between the X-ray and VHE flux, and more notably direct correlation without any lead or lag time, is a natural signature of SSC emission.  Within the single-zone SSC paradigm, the inverse-Compton flux is emerging from the same region as the synchrotron emission, and is fundamentally derived from the same particle and photon populations as the synchrotron emission.  In this way, any variability in the synchrotron photon luminosity will immediately be translated into a change in the up-scattered inverse Compton luminosity.

In applying a single-zone equilibrium SSC model to the broadband data of Mrk 501, we find that the data could be reasonably represented in each of the five simultaneous epochs.  Notably, the injected particle populations on MJD 56485.0, 56485.9 and 56486.9 are very hard, with an injection index of $q=1.3$.  Such a hard injection index is difficult to produce with standard shock acceleration scenarios alone, but is possible through a magnetic reconnection event (e.g. as explained in  \citealt{romanova, sironi, guo}).  The increase in energy of the particle population (with an additional hardening to the injection index of $q=1.3$) between the SED derived for MJD 56485.0 as compared to MJD 56420 indicates an introduction of additional energetic particles to the emission region, requiring some source of energy input.  The decrease of the magnetic field, similar to what would naturally occur after a magnetic reconnection event, is capable of accelerating particles near the point of reconnection and producing the newly injected $q=1.3$ particle population.  Additionally, the decrease in the emission region size is consistent with a magnetic reconnection event that affects a more localized region as compared to a larger, more steady non-thermal emission region.   More information on particle acceleration via magnetic reconnection can be found in \cite{werner} and \cite{guo2015}.


The variability timescale for these model representations, quoted in Table 6, is determined from the light-crossing timescale of the emission region according to $t_{\rm var}=R/c \delta (1+z)$.    For the emission region sizes and Doppler factor of $\delta$=15 used within the model, the predicted variability timescales of a couple of hours are compatible with the variability timescale observed during the broadband observations.  The radiative cooling timescale is approximately equal to the synchrotron cooling timescale, $t_{\rm sync}\sim 1.4\times10^{4}\,(B_0 / 0.06 \,\rm G)^{-2} \gamma_6^{-1}$\,s, where $\gamma_6=\gamma/(10^6)$.  With a minimum light crossing time, corresponding to the minimum variability timescale of $t_{\rm var}\sim1.6\times10^4$\,s (in the observer frame), all but the most energetic electrons within the emitting region cool on timescales that are longer than the crossing timescale, showing that the observed variability is likely a reflection of changes in the particle acceleration and/or injection processes directly.

Notably, faster variability timescales have been observed from Mrk 501 in the past \citep[e.g.][]{mrk501MAGIC1} and so the model parameters shown here cannot be generalized to all Mrk 501 flux variability episodes.   \textit{NuSTAR} observations show the hard X-ray flux to significantly decrease by more than $10\%$ between its 90-min orbits.   Moreover, on MJD 56420 the source hard X-ray flux was observed to change by a factor of greater than 40\% in the 7-30 keV band during a 7 hour exposure.  

In an attempt to describe a possible emission scenario which might result in the broadband SED variability observed for Mrk 501 in 2013, the parameter changes were made to the single zone equilibrium SSC model monotonically.  With a degeneracy between several of the input parameters, the model applied here cannot be used for conclusive studies regarding which changes occur within the emitting region from one state to the next.  Instead, through the study of band-to-band spectral variability, leads and/or lags and fractional variability, as well as broadband modeling of various flaring episodes, we find compelling evidence to support a single zone SSC emission scenario for Mrk 501 during the broadband observations in this campaign.

The collection of simultaneous broadband observations is a necessity for the study of the relativistic emission mechanisms at work within blazars such as Mrk 501.  It is known that these sources vary continually, with characteristics that significantly change between different flaring episodes, requiring the continuation of deep broadband observations such as those presented in this work.

\acknowledgments
This work was supported under NASA Contract No. NNG08FD60C, and  made use of data from the \textit{NuSTAR} mission, a project led by the California Institute of Technology, managed by the Jet Propulsion Laboratory, and funded by the National Aeronautics and Space Administration. We thank the \textit{NuSTAR} Operations, Software and Calibration teams for support with the execution and analysis of these observations. ÊThis research has made use of the \textit{NuSTAR} Data Analysis Software (NuSTARDAS) jointly developed by the ASI Science Data Center (ASDC, Italy) and the California Institute of Technology (USA).

The MAGIC Collaboration would like to thank
the Instituto de Astrof\'{\i}sica de Canarias
for the excellent working conditions
at the Observatorio del Roque de los Muchachos in La Palma.
The financial support of the German BMBF and MPG,
the Italian INFN and INAF,
the Swiss National Fund SNF,
the ERDF under the Spanish MINECO, and
the Japanese JSPS and MEXT
is gratefully acknowledged.
This work was also supported
by the Centro de Excelencia Severo Ochoa SEV-2012-0234, CPAN CSD2007-00042, and MultiDark CSD2009-00064 projects of the Spanish Consolider-Ingenio 2010 programme,
by grant 268740 of the Academy of Finland,
by the Croatian Science Foundation (HrZZ) Project 09/176 and the University of Rijeka Project 13.12.1.3.02,
by the DFG Collaborative Research Centers SFB823/C4 and SFB876/C3,
and by the Polish MNiSzW grant 745/N-HESS-MAGIC/2010/0.

This research is also supported by grants from the U.S. Department of Energy Office of Science, the U.S. National Science Foundation and the Smithsonian Institution, by NSERC in Canada, by Science Foundation Ireland (SFI 10/RFP/AST2748) and by STFC in the U.K. We acknowledge the excellent work of the technical support staff at the Fred Lawrence Whipple Observatory and at the collaborating institutions in the construction and operation of the VERITAS instrument.  The VERITAS Collaboration is grateful to Trevor Weekes for his seminal contributions and leadership in the field of VHE gamma-ray astrophysics, which made this study possible.

The \textit{Fermi} LAT Collaboration acknowledges generous ongoing support
from a number of agencies and institutes that have supported both the
development and the operation of the LAT as well as scientific data analysis.
These include the National Aeronautics and Space Administration and the
Department of Energy in the United States, the Commissariat \`a l'Energie Atomique
and the Centre National de la Recherche Scientifique / Institut National de Physique
Nucl\'eaire et de Physique des Particules in France, the Agenzia Spaziale Italiana
and the Istituto Nazionale di Fisica Nucleare in Italy, the Ministry of Education,
Culture, Sports, Science and Technology (MEXT), High Energy Accelerator Research
Organization (KEK) and Japan Aerospace Exploration Agency (JAXA) in Japan, and
the K.~A.~Wallenberg Foundation, the Swedish Research Council and the
Swedish National Space Board in Sweden.
 
Additional support for science analysis during the operations phase is gratefully acknowledged from the Istituto Nazionale di Astrofisica in Italy and the Centre National d'\'Etudes Spatiales in France.

We thank the \textit{Swift} team duty scientists and science planners and we acknowledge the use of public data from the $Swift$ data archive. This research has made use of the $Swift$ XRT Data Analysis Software (XRTDAS) developed under the responsibility of the ASI Science Data Center (ASDC), Italy.

This research is partly based on observations with the 100-m telescope of
the MPIfR (Max-Planck-Institut f\"ur Radioastronomie) at Effelsberg and with the IRAM 30-m
telescope. IRAM is supported by INSU/CNRS (France), MPG (Germany) and IGN
(Spain). V. Karamanavis and I. Myserlis are funded by the International Max Planck Research
School (IMPRS) for Astronomy and Astrophysics at the Universities of Bonn and Cologne.

The OVRO 40-m monitoring program is supported in part by NASA grants NNX08AW31G
and NNX11A043G, and NSF grants AST-0808050  and AST-1109911.

TG acknowledges support from Istanbul University (Project numbers
49429 and 48285),  Bilim Akademisi (BAGEP  program) and TUBITAK
(project numbers 13AT100-431, 13AT100-466, and 13AT60-430).

St.Petersburg University team acknowledges support from Russian RFBR
grant 15-02-00949 and St. Petersburg University research
grant 6.38.335.2015.

The research at Boston University (BU) was funded in part by NASA Fermi Guest Investigator grant NNX14AQ58G
and Swift Guest Investigator grant NNX14AI96G. The PRISM camera at Lowell Observatory was developed by K.\ Janes et al. at BU and Lowell Observatory, with funding from the NSF, BU, and Lowell Observatory.

This research was partially supported by Scientific Research Fund of the
Bulgarian Ministry of Education and Sciences under grant DO 02-137
(BIn-13/09). The Skinakas Observatory is a collaborative project of the
University of Crete, the Foundation for Research and Technology -- Hellas,
and the Max-Planck-Institut f\"ur Extraterrestrische Physik.

The Abastumani Observatory team acknowledges financial support by the by
Shota Rustaveli National Science Foundation under contract FR/577/6-320/13.

We are grateful to the American Association of Variable Star Observers (AAVSO) for assisting us with some of the optical data acquisition.

G.Damljanovic and O.Vince gratefully acknowledge the observing grant support from the Institute
of Astronomy and Rozhen National Astronomical Observatory, Bulgarian
Academy of Sciences. This work is in accordance with the Projects
No 176011 (``Dynamics and kinematics of celestial bodies and systems"),
No 176004 (``Stellar physics") and No 176021 (``Visible and invisible
matter in nearby galaxies: theory and observations") supported by the
Ministry of Education, Science and Technological Development of
the Republic of Serbia.

{\it Facilities:} \facility{NuSTAR}, \facility{MAGIC}, \facility{VERITAS}, \facility{Fermi LAT}, \facility{Swift XRT/UVOT}, \facility{West Mountain Observatory (WMO)}, \facility{AAVSOnet}, \facility{FLWO}, \facility{TUBITAK}, \facility{GASP}, \facility{St. Petersburg Crimean}, \facility{Perkins}, \facility{Lulin}, \facility{Rozhen}, \facility{Skinakas}, \facility{Abastumani}, \facility{OVRO}, \facility{Metsahovi}.

\end{document}

%% file: SwiftXRT.tex
\begin{deluxetable*}{cccccccccccc}
\tabletypesize{\scriptsize}
\tablecaption{\textit{Swift} XRT observations and analysis results for \textit{NuSTAR}-simultaneous periods.  Integral flux values are calculated according to the PL model, and are provided in $\times10^{-11}$ erg cm$^{-2}$ s$^{-1}$ units. The errors for each parameter are found using a value of $\Delta\chi^2$=2.706, corresponding to a 90\% confidence level for a parameter. }
\tablewidth{0pt}
\tablehead{
  \colhead{Observation}&  \colhead{Date} &\colhead{Exp}&  \colhead{Flux} &     \colhead{Flux}& \colhead{Flux} &  \colhead{Flux}&  \colhead{Index}&   \colhead{$\chi^2$/DOF}& \colhead{$\Gamma$}& \colhead{$\beta$} & \colhead{$\chi^2$/DOF}\\
  \colhead{ID}&  \colhead{[MJD]}& \colhead{[s]}& \colhead{2-10 keV}&  \colhead{0.5-2 keV} &    \colhead{3-7 keV}&  \colhead{0.3-3 keV} & \colhead{$\Gamma$}&  \colhead{}&  \colhead{LP}&  \colhead{LP} &\colhead{}  \\
}
 \startdata
00080176001 & 56395.06  & 9636.0 & 6.9$\pm$0.1 & 6.41$\pm$0.06 & 3.6$\pm$0.1 & 11.0$\pm$0.1 & 2.05$\pm$0.01 & 403.5/416 & 2.06$\pm$0.02 & -0.02$\pm$0.04 & 402.6/415 \\
00091745001 & 56485.84  & 250.7 & 21.1$\pm$1.7 & 12.7$\pm$0.4 & 10.9$\pm$0.9 & 22.3$\pm$0.7 & 1.77$\pm$0.05 & 108.1/94 & 1.74$\pm$0.08 & 0.10$\pm$0.16 & 107.0/93 \\ 
00030793235 & 56485.98  & 709.1 & 24.3$\pm$1.1 & 14.6$\pm$0.2 & 13.1$\pm$0.9 & 24.1$\pm$0.4& 1.77$\pm$0.03 & 228.7/222 & 1.75$\pm$0.05 & 0.03$\pm$0.09 & 227.6/221 \\
00030793236 & 56486.31  & 1002.0 & 24.0$\pm$0.7 & 14.1$\pm$0.3 & 13.4$\pm$0.6 & 23.4$\pm$0.4 & 1.73$\pm$0.03 & 291.6/270 & 1.68$\pm$0.04 & 0.13$\pm$0.08 & 285.1/269 \\ 
00030793237 & 56487.04  & 949.5 & 19.1$\pm$0.9 & 12.0$\pm$0.2 & 10.4$\pm$0.4 & 18.9$\pm$0.3 & 1.76$\pm$0.03 & 229.9/237 & 1.73$\pm$0.05 & 0.07$\pm$0.08 & 228.9/236 \\ 
 \enddata
\end{deluxetable*}